\documentclass[twocolumn]{aastex63}

\usepackage{xcolor}
\usepackage[caption=false]{subfig}
\usepackage[normalem]{ulem}
\usepackage[flushleft]{threeparttable}

\newcommand{\E}[1]{\times 10^{#1}}

\newcommand{\s}{\,{\rm s}}      
\newcommand{\ps}{\,{\rm s}^{-1}}
    
\newcommand{\Msun}{M_{\odot}}
\newcommand{\cm}{\,{\rm cm}}    
\newcommand{\km}{\,{\rm km}}

\newcommand{\K}{\,{\rm K}}

\newcommand{\g}{\,{\rm g}}

\newcommand{\nel}{n_{e}}        
\newcommand{\NH}{N_{\rm H}}

\newcommand{\XMMN}{{\sl XMM-Newton}}
\newcommand{\Suzaku}{{\sl Suzaku}}

\newcommand{\Chandra}{{\sl Chandra}}

\newcommand{\snr}{Sgr A East}

\received{July 29, 2020}
\accepted{September 8, 2020}
\shorttitle{Sgr A East: evidence for a Type Iax SNR}
\shortauthors{Zhou et al.}
\graphicspath{{./}{figures/}}

\begin{document}

\title{
Chemical abundances in Sgr A East: evidence for a Type Iax supernova remnant
}

\author{Ping Zhou}
\affil{Anton Pannekoek Institute for Astronomy, University of Amsterdam, Science Park 904, 1098 XH Amsterdam, The Netherlands}
\email{p.zhou@uva.nl}
\affil{School of Astronomy and Space Science, Nanjing University, Nanjing 210023, PR China}

\author{Shing-Chi Leung}
\affil{TAPIR, Walter Burke Institute for Theoretical Physics, Mailcode 350-17, Caltech, Pasadena, CA 91125, USA}
\email{scleung@caltech.edu}

\author{Zhiyuan Li}
\affil{School of Astronomy and Space Science, Nanjing University, Nanjing 210023, PR China}
\affil{Key Laboratory of Modern Astronomy and Astrophysics, Nanjing University, Ministry of Education, PR China}

\author{Ken'ichi Nomoto}
\affil{Kavli Institute for the Physics and Mathematics of the Universe (WPI), The University of Tokyo, Kashiwa, Chiba 277-8583, Japan}

\author{Jacco Vink}
\affil{Anton Pannekoek Institute for Astronomy, University of Amsterdam, Science Park 904, 1098 XH Amsterdam, The Netherlands}
\affil{GRAPPA, University of Amsterdam, PO Box 94249, 1090 Amsterdam, The Netherlands}
\affil{SRON, Netherlands Institute for Space Research, Sorbonnelaan 2, 3584 Utrecht, The Netherlands}

\author{Yang Chen}
\affil{School of Astronomy and Space Science, Nanjing University, Nanjing 210023, PR China}
\affil{Key Laboratory of Modern Astronomy and Astrophysics, Nanjing University, Ministry of Education, PR China}



\begin{abstract}

Recent observations have shown a remarkable diversity of observational behaviors and explosion mechanisms in thermonuclear supernovae (SNe).
An emerging class of peculiar thermonuclear SNe, called Type Iax, show photometric and spectroscopic behaviors distinct from normal Type Ia.
Their origin remains highly controversial, but pure turbulent deflagration of white dwarfs (WDs) has been regarded as the leading formation theory.
The large population of Type Iax indicates the existence of unidentified Galactic Type Iax supernova remnants (SNRs).
We report evidence that  
SNR \snr\ in the Galactic center
resulted from a pure turbulent deflagration of a Chandrasekhar-mass carbon--oxygen WD, an explosion mechanism used for Type Iax SNe.
Our X-ray spectroscopic study of \snr\ using 3~Ms of Chandra data shows a low ratio of intermediate-mass elements to Fe and large Mn/Fe and Ni/Fe ratios.
This abundance pattern does not accord with the 
core-collapse or normal Type Ia models.
Sgr A East is thus the first Galactic SNR for which a likely Type Iax origin has been proposed and the nearest target 
for studying this peculiar class. 
We compared \snr\ with the Fe-rich SNRs 3C~397 and W49B, which
also have high Mn and Cr abundances and were claimed to result from deflagration-to-detonation explosions of Chandrasekhar-mass WDs (although with disputes).
Our study shows that they have distinct abundance patterns.
The X-ray spectroscopic studies of thermonuclear SNRs provide observational evidence for the theories that there are diverse explosion channels and various metal outputs for Chandrasekhar-mass WDs.

\end{abstract}



\section{Introduction} \label{sec:intro}

Thermonuclear supernovae (SNe) are factories of iron-group elements (IGEs, such as Fe, Ni, Mn, Cr) in
our universe, but their metal yields are sensitive to the explosion mechanisms \citep[e.g.,][]{seitenzahl17}.
Recent observations have shown that the thermonuclear SN zoo includes more than
Type Ia SNe
\citep[see][for a recent review]{jha19}.
The observed diversity provides opportunities to study different explosion mechanisms of white dwarfs (WDs).

Type Iax SNe are the largest class of peculiar  thermonuclear SNe with 
observational behaviors similar to SN~2002cx, which was initially considered as the most peculiar Type Ia 
SN \citep{li03}.
They are distinguished from normal Type Ia because they
show
lower luminosities and ejecta 
velocities and masses, implying that the Type Iax and Type Ia classes are created from different explosion mechanisms \citep{foley13}.

The current understanding of Type Iax SNe has only been obtained from extragalactic SNe, lacking observations of detailed chemical composition in SN ejecta.
Their origin has been highly controversial \citep[e.g.,][]{lyman13,mccully14} and their later evolution has not been observed in detail.
Among those controversial models for Type Iax SNe, 
promising are the pure deflagration models of a WD with a bound remnant \citep[see][for a review]{jha17},
which predict that the ejecta has large Mn, Cr, and Ni to Fe ratios, but a low ratio of intermediate-mass elements (IMEs) to Fe \citep{fink14,leung20Iax}.

\begin{figure*}
\epsscale{0.8}
\plotone{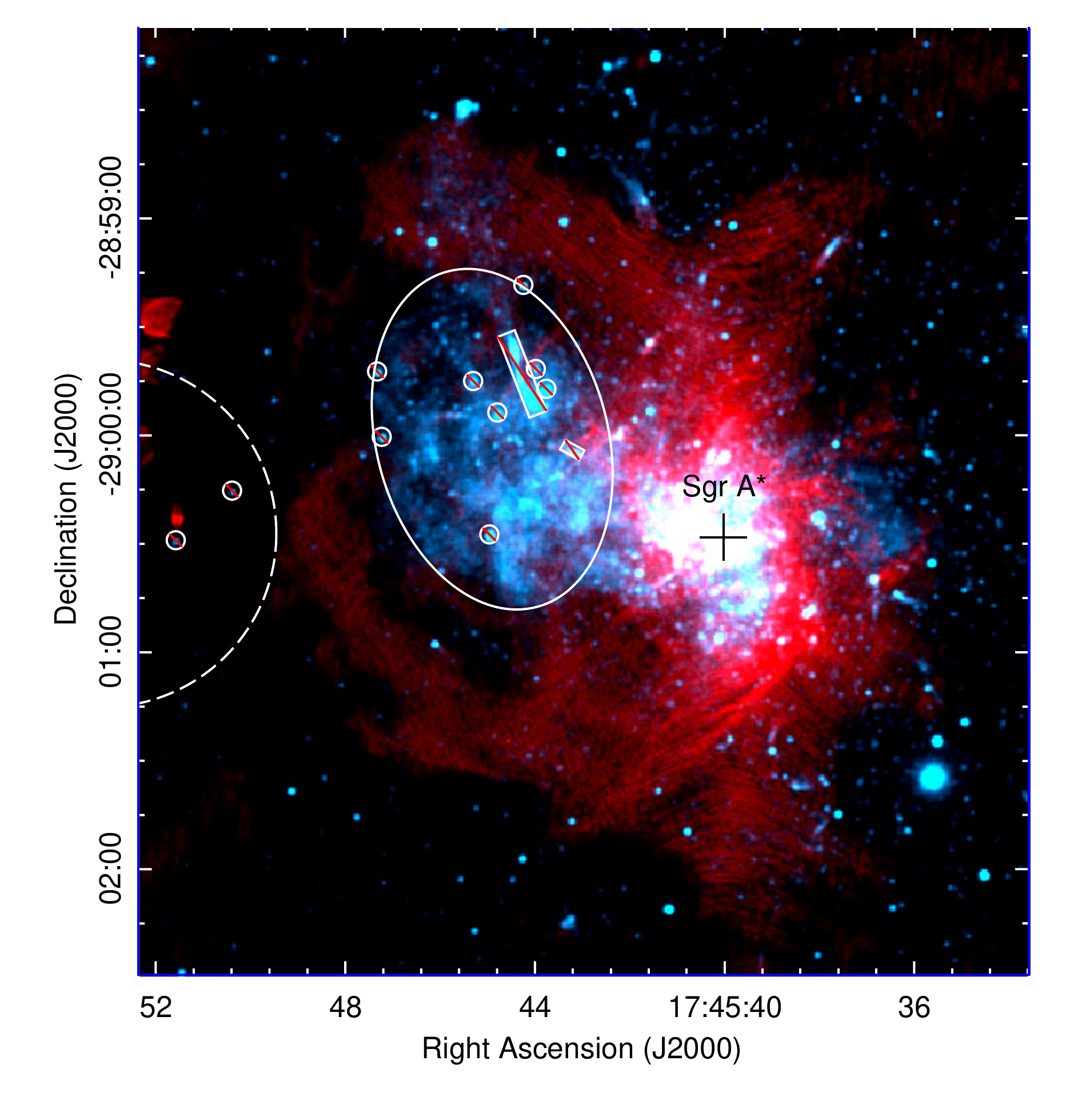}
\caption{
Composite image of \snr.
Red:VLA 8.3~GHz image \citep{zhao09};
Cyan: {\it Chandra} 2--8~keV image.
The spectral extraction region
(solid) and background region (dashed) are overlaid.
The small circles and rectangles denote bright point sources and nonthermal filaments that are probably interlopers and thus masked from spectral extraction. 
}
\label{fig:x_c}
\end{figure*}

The population of Type Iax is found to be large, with 2--5 for every 10 Type Ia \citep{foley13}. 
Finding a remnant of a nearby Type Iax SN would provide essential insight into this group.
However,
none of 294--383 known Galactic supernova remnants \citep[SNRs;][]{green17,ferrand12} has been reported to
have a Type Iax origin.
The Type Ia SN rate of $19\pm 6\%$ \citep{li11} could be translated to $\sim 40$--100 Type Ia SNRs in our Galaxy.
Given the large occurrence rate of Type Iax SNe, we roughly estimate that 
a few to $\sim 50$ Type Iax SNRs in our Galaxy are waiting to be identified.

\snr\ (G0.0+0.0) is the only known 
SNR strikingly close to the Galactic-center super-massive black hole,
originally identified from radio observations \citep{eker83}.
A few arguments support that  \snr\ is interacting 
with the  molecular ridge  
in the central parsecs
and is likely overrunning the circumnuclear disk orbiting the supermassive black hole Sgr A*  
\citep[see Figure \ref{fig:x_c} and][]{rockefeller05}.
\snr\ has been presumed to result from
a core-collapse SN explosion of a $<20~M_\odot$ star \citep{maeda02,park05}, although 
a Type Ia origin has not been ruled out, as Fe is found to be more abundant than IMEs in the remnant \citep{sakano04,park05}.
\cite{one19} reported Mn and Cr lines from \snr\ using \Suzaku\ data.
Mn and Cr are more frequently observed in thermonuclear SNRs
\citep[see][and references therein]{yang13}, although they have also been found in the X-ray-luminous core-collapse SNR Cassiopeia A  
\citep[e.g.,][]{sato20b}.

SNR origin can often be probed through metal composition in the ejecta as diagnosed by X-ray spectroscopy
\citep{vink12}.
We utilize the 3~million second (Ms) {\it Chandra} X-ray data taken in 2012, with a total exposure    
more than five times that of any previous X-ray studies.
The deep observations allow us to clearly detect lines such as Mn, Cr, and Fe 
and use the metal composition of the remnant to unveil its SN origin.

The paper is organized as follows.
Section~2 describes the Chandra X-ray data used for this study.
The analysis of the X-ray data is presented in Section~3.
In Section 4, we compare the abundance pattern of \snr\ with 
various SN nucleosynthesis models and discuss
its Type Iax origin. This section also includes a discussion about \snr\ has distinguished abundances from 3C~397 and W49B, two middle-aged SNRs that also show high abundances of Mn, Cr, and Fe and were proposed to have a thermonuclear SN origin \citep[][]{yamaguchi15,zhou18a}.

\section{Chandra X-ray Data}
The inner parsecs of the Galactic center, where Sgr A East is located, have been frequently observed by the {\it Chandra} X-ray observatory, primarily with its Advanced CCD Imaging Spectrometer (ACIS). 
In this work, we utilized a total of 38 observations taken in 2012 for spectroscopic analysis. These observations were taken with the combined operation of ACIS-S and the High Energy Transmission Grating (HETG), in a total exposure of 2.94 Ms. With the HETG inserted, about one half of the incident X-rays are dispersed, while the remaining X-rays continue to the detector directly and form the ``zeroth-order" image. We used only data from the ``zeroth-order" image (here referred to as the ACIS-S image for simplicity). While there also exist a large number of ACIS-I observations toward the Galactic center, we do not include these data for two reasons: (1) in most cases, Sgr A East falls on the CCD gaps of the ACIS-I array, potentially introducing systematic uncertainty in the instrumental response; (2) it is known that charge-transfer inefficiency degrades the spectral resolution of ACIS, in the sense that signals recorded in larger detector rows have a poorer resolution. Therefore, at the position of Sgr A East, the ACIS-S observations have a better spectral resolution than that of the ACIS-I, which is desirable for accurate measurement of emission lines expected from the hot plasma.
To maximize the X-ray photons shown in Figure~\ref{fig:x_c}, we combined the ACIS-S image taken in 2012 and the ACIS-I image taken during 1999 and 2013. 
The detailed observation information is listed in Table~1 in \cite{zhu18}. 

We uniformly reprocessed the archival data with CIAO v4.10 and calibration files CALDB v4.7.8, following the procedures detailed in \cite{zhu19}.
We have examined the light curve of each ObsID and, when necessary, removed time intervals contaminated by particle flares. 
We use XSPEC \citep[vers 12.10.1f]{arnaud96} with ATOMDB 3.0.9
\footnote{http://www.atomdb.org/} 
for spectral analysis.

\section{Spectral analysis}

We extracted the global spectrum from the X-ray-bright region in the SNR interior 
(the solid ellipse with a short half-axis of $32''$ and a long half-axis of $48''$, in Figure~1) and subtracted the background
from a source-free region outside the SNR boundary (the dashed circle).
We removed several bright point-like sources with 
an observed 2-8 keV photon flux greater than $8\times10^{-7}{\rm~photons~cm^{-2}~s^{-1}}$
and two nonthermal filaments.
We coadded the 38 individual spectra to produce a combined spectrum of high signal-to-noise ratio, weighting the corresponding ancillary response files (ARFs) and redistribution matrix files (RMFs) by the effective exposure.

\begin{figure}
\epsscale{1.2}
\plotone{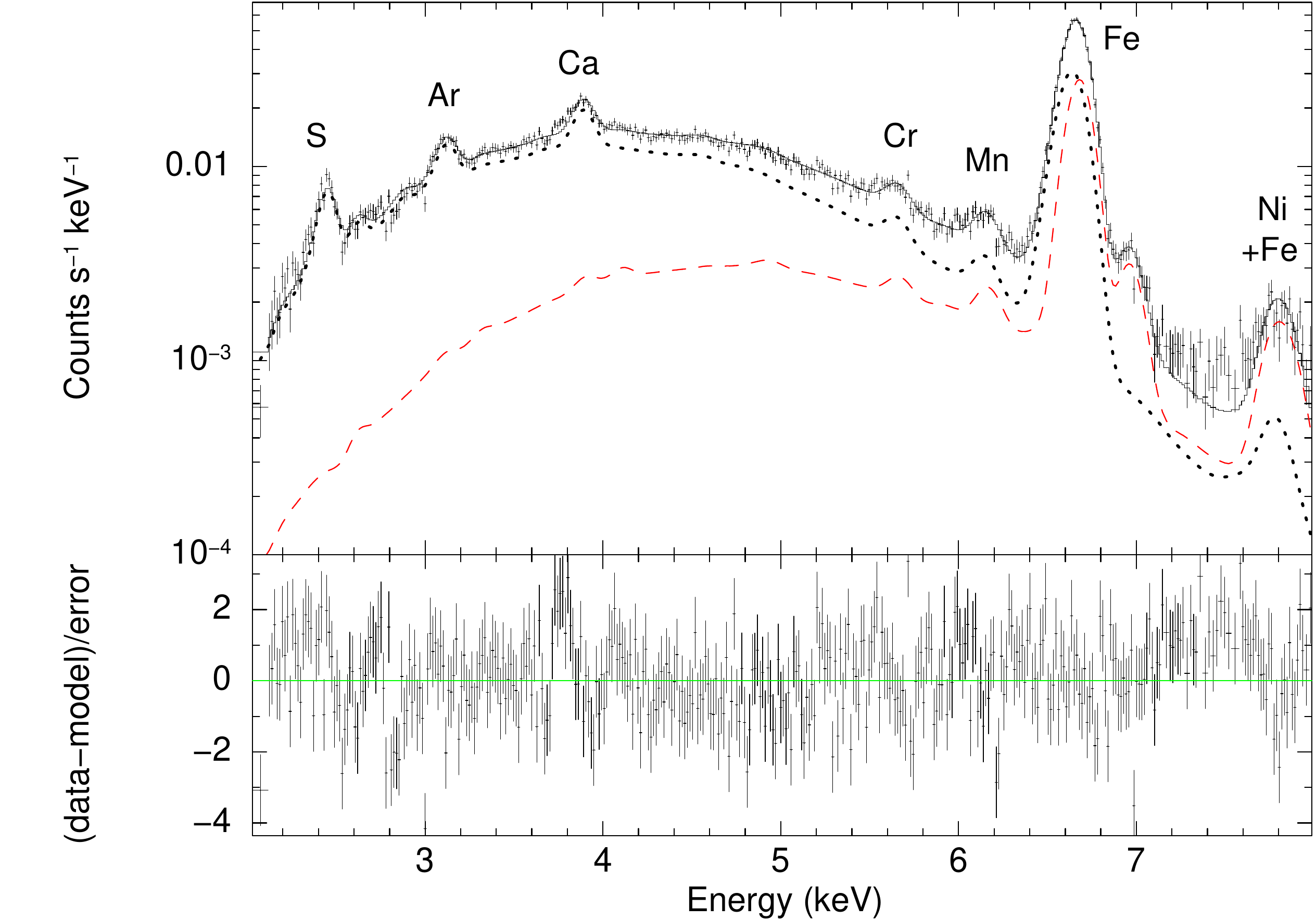}
\caption{
{\it Chandra} ACIS-S spectrum in 2--8 keV fitted 
with an absorbed two-temperature model $tbabs*(vvapec+vvapec)$.
The dotted line and the red dashed line denote
the cool and hot components, respectively.
Spectral fit results are shown in Table~\ref{tab:fit}.
}
\label{fig:spec}
\end{figure}

Figure~\ref{fig:spec} shows the ACIS S3-chip spectrum of \snr\ extracted
from the X-ray-bright interior (see Figure~\ref{fig:x_c}).
The combined spectrum of \snr\ in 2--8 keV shows emission
lines of S, Ar, Ca, Fe, and Ni.
We confirm that Cr and Mn  He$\alpha$ lines are detected at 5.63~keV and 6.13~keV, respectively \citep{one19}.
The Fe~He$\alpha$ and Ly$\alpha$ lines are also separated using the ACIS-S chips,
while the ACIS-I cannot resolve them \citep[see ACIS-I spectra in][]{park05}.

We first summarize the best-fit results of our spectral analysis, followed by a detailed description of the spectral fit in
Sections~\ref{sec:singlecomp} and \ref{sec:doublecomp}. 
We found that a two-temperature plasma model can
well describe the X-ray spectrum, providing 
a reduced $\chi^2$ of $\chi_\nu^2=1.45$ with a degree of freedom (dof) of 381 (see Figure~\ref{fig:spec}), while none of the single thermal
models can fit the spectrum well ($\chi_\nu^2 \gtrsim 2$, see Table~\ref{tab:fit}).
The X-ray emission is characterized by a cool
component with a temperature $kT_{\rm c}=1.19\pm 0.03$~keV
and a hot component with a temperature $kT_{\rm h}=4.3\pm0.2$~keV, with a foreground
absorption of $\NH=2.14\pm 0.04 \E{23}~\cm^{-2}$.
Both components are in (near-)collisional ionization equilibrium (CIE).
The two temperatures are consistent with those suggested in previous X-ray studies of \snr\ \citep{sakano04,park05}.
Moreover, we found very high abundances of IGEs 
Cr ($5.6\pm 1.0$), Mn ($17\pm 3$), Fe 
($6.5\pm 0.4$), and Ni ($17\pm4$) 
compared with solar abundances.
Quoted errors here are at a 90\% confidence level. 
The uncertainty of the Ni abundance could be larger than the fitted value as residuals ($\lesssim2\sigma$) are shown in 7--8~keV.
In contrast, the IMEs  S, 
Ar, and Ca  have ordinary abundances ($1.4\pm 0.2$, $1.5\pm 0.2$ 
and $1.7\pm 0.1$, respectively), which are similar to the mean interstellar medium values of $\sim 2$ in the Galactic center \citep{mezger96}, 
and consistent with the S and Ca abundances obtained from the observations of HII regions \citep[$Z_{\rm S}/Z_{\rm S}^\odot\sim 1.5$,][]{rudolph06} and red supergiants \citep[$Z_{\rm Ca}/Z_{\rm Ca}^\odot\sim 1.7$,][]{davies09}.

\begin{table*}
\footnotesize
\centering
\caption{Best-fit results of the global spectrum with 90\% uncertainties \label{tab:fit}}
\begin{threeparttable}
\begin{tabular}{lcccccc}
\hline
\hline
Energy Range & \multicolumn2c{2--8 keV} & &
\multicolumn3c{5--8 keV} \\  \cline{2-3} \cline{5-7} 
Model &  $vvnei+vvapec$ & $vvapec+vvapec$ & 
& $vvapec$ & $vvrnei$ & $bvvrnei^a$ \\
\hline
$\chi_\nu^2$/dof & 1.43/380 & 1.45/381 
& & 3.94/188 & 3.00/186  & 1.95/185\\
$N_{\rm H} (10^{23}$~cm$^{-2})$  &
$2.07\pm 0.05$ & $2.14\pm 0.04$  &
& 2.14 (fixed) & 2.14 (fixed) & 2.14 (fixed)\\
$kT_{\rm c}$ (keV) &  
$1.22\pm 0.04$ & $1.19\pm 0.03$ 
& & $\sim 1.62$ & $\sim 1.52$ & $1.52\pm 0.03$\\
$kT_{\rm c}^{\rm init}$ (keV) & 
\dots  & \dots 
& &\dots  & $\sim 11.7$ & $> 7.7$ \\
$\tau_{\rm c}~(10^{11}$~cm$^{-3}$ s)$^b$
& $>7.2$ &\dots  & 
& \dots & $\sim 8.8$  & $8.5^{+1.8}_{-1.4}$\\
$kT_{\rm h}$ (keV) 
& $4.0\pm 0.2$ &  $4.3\pm 0.2$ 
&  & \dots & \dots & \dots\\
$norm_{\rm c} (\times 10^{-2})$
& $4.4\pm 0.4$
& $4.8\pm 0.3$
&  & $\sim 3.4$ 
& $\sim 3.8$ & $3.9\pm 0.2$\\
$norm_{\rm h} (\times 10^{-3})$  
& $2.6\pm 0.3$ & $2.0\pm 0.3$
& & \dots & \dots & \dots
\\
\hline
S $^c$ 
& $1.4\pm 0.2$ & $1.4\pm 0.2$ 
& & \dots & \dots & \dots
\\
Ar 
& $1.5\pm 0.2$ & $1.5\pm 0.2$ 
& & \dots & \dots & \dots
\\
Ca 
& $1.7\pm 0.1$ & $1.7\pm 0.1$ 
& & \dots & \dots & \dots
\\
Cr 
& $4.7\pm 1.0$  & $5.6\pm 1.0$ 
& & $\sim 2.7$ & $\sim 3.6$ & $3.0\pm 0.6$ \\
Mn 
& $14.5\pm 2.5$ & $16.9 \pm 2.6$ 
& & $\sim 8.7$ & $\sim 10.7$ & $9.0\pm 1.2$\\
Fe 
& $6.2\pm 0.3$ & $6.5 \pm 0.4$ 
& & $\sim 4.3$ & $\sim 4.0$ & $3.5\pm 0.2$\\
Ni 
& $14.6\pm 3.1$ & $17.0 \pm 3.5$ 
& & $\sim 20$ & $\sim 17$ & $12.4\pm 2.3$ \\
\hline
\hline
\end{tabular}
\begin{tablenotes}
\item $^a$ In the $bvvrnei$ model, the best-fit Gaussian sigma for the velocity broadening is $1.44\pm 0.09\E{3}\km\ps$.
\item $^b$ The ionization timescale $\tau_c=\int \nel dt\sim \nel t$ for the underionized plasma model $vvnei$ model or
recombining timescale for the recombining plasma $bvvrnei$ model.
It describes an electron density $\nel$-weighted timescale, starting from an instantaneous heating ($vvnei$) or cooling ($vvrnei$).
\item $^c$ The abundance ratio of the element relative to its solar value (e.g, for S, the value means $Z_{\rm S}/Z_{\rm S}^\odot$).
\end{tablenotes}
\end{threeparttable}
\end{table*} 

\subsection{Single-component model} \label{sec:singlecomp}
\begin{figure*}
  \centering
  \includegraphics[angle=0, width=0.8\textwidth]{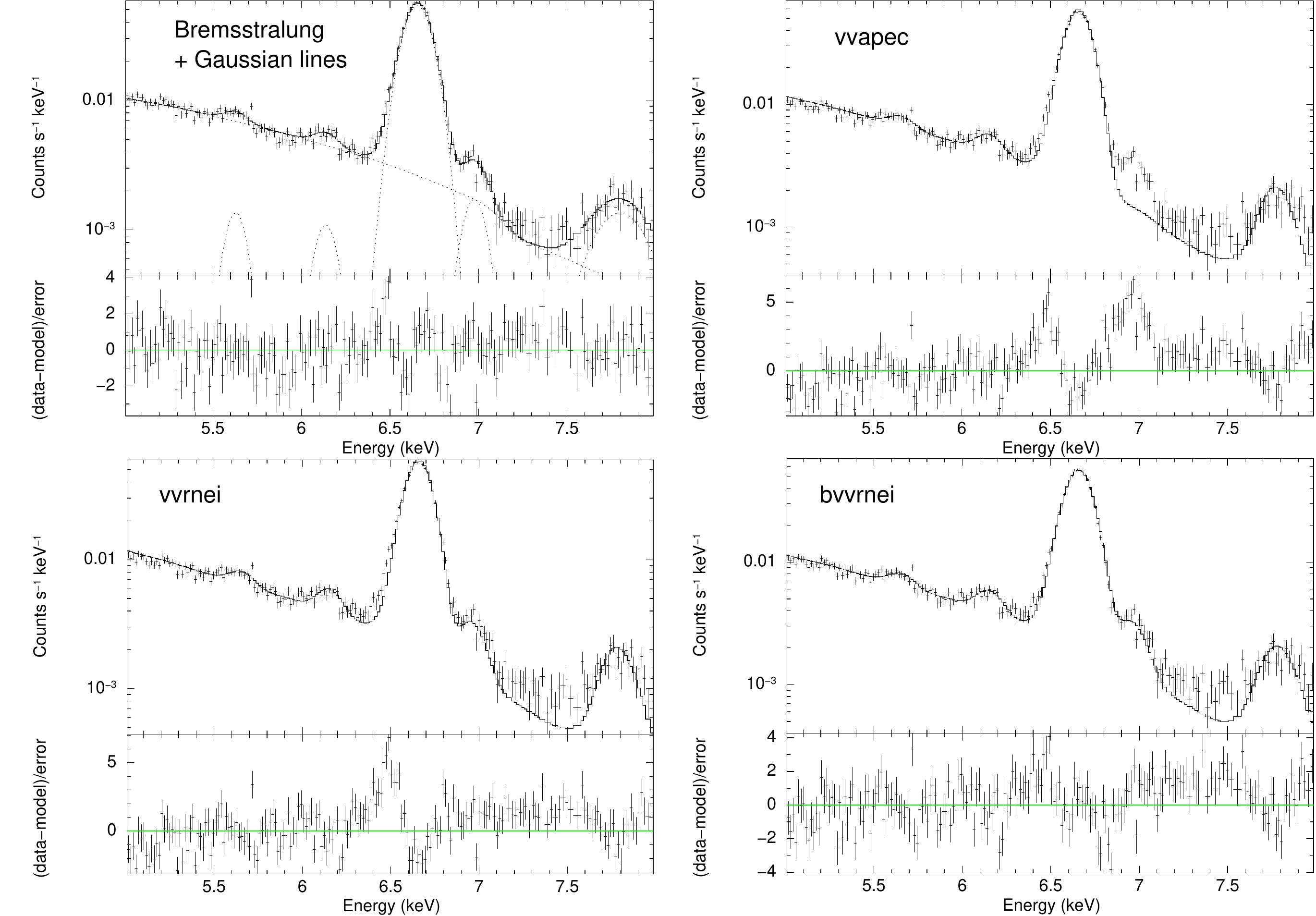}
\caption{
{\it Chandra} ACIS-S spectrum in 5--8~keV fitted with single thermal component models. Top left: the spectrum in 5--8~keV
fitted using absorbed bremsstrahlung$+$Gaussian lines (see Table~\ref{tab:line} for the fitting results);
top right: $vvapec$; bottom left: recombining plasma model $vvrnei$; and bottom right: recombining plasma model $bvvrnei$ with a velocity broadening.
}
\label{fig:4spec}
\end{figure*}

We first investigated the hard X-ray emission in 5--8 keV, where
IGE (Cr, Mn, Fe, and Ni) lines are shown. 
As the hard X-ray spectrum cannot well constrain $\NH$, we fixed
$\NH$ value to $2.14\E{23}~\cm^{-2}$, the best-fit result from fitting 
the 2--8~keV spectrum with a two-temperature model.
Table~\ref{tab:line} shows the
line properties of Cr, Mn, and Fe,
which are obtained by simply fitting the 5--8~keV spectrum with an absorbed bremsstrahlung component plus Gaussian lines
(see Figure~\ref{fig:4spec}).
The photon fluxes for the Cr and Mn lines are around
$3\times 10^{-6}$~photon~s$^{-1}$ cm$^{-2}$.
The Fe Ly$\alpha$ to He$\alpha$ intensity ratio of
3\% implies an ionization temperature of 
$\sim 3.0$~keV (ATOMDB 3.0.9), which 
is higher than the 
electron temperature of $2.0\pm 0.1$~keV obtained from the bremsstrahlung continuum.
This implies that either 
the gas is overionized or
the X-ray plasmas have
multitemperature components.

\begin{deluxetable}{lcccc|lll}
\tabletypesize{\scriptsize}
\tablewidth{0pt} 
\tablecaption{Line properties for the Fe-group elements and the
90\% uncertainties \label{tab:line}}
\tablehead{
\colhead{Line} & \colhead{Central Energy}
& \colhead{Width$^a$}  &  \colhead{Photon Flux} \\
& \colhead{(keV)} & \colhead{(eV)} &  \colhead{(photons s$^{-1}$ cm$^{-2}$)} }
\startdata 
Cr He$\alpha$ & $5.63\pm 0.02$ &  \dots & $3.2\pm 0.8 \times 10^{-6}$ \\
Mn He$\alpha$ & $6.13\pm 0.02$ &  \dots & $2.8\pm 0.7 \times 10^{-6}$ \\
Fe He$\alpha$ & $6.654\pm 0.001$ &  $45\pm1 $ & $2.03\pm 0.02\times 10^{-4}$ \\
Fe Ly$\alpha$ & $6.978\pm 0.003$ &  \dots & $6.6\pm 0.9 \times 10^{-6}$ \\
Ni He$\alpha$+Fe lines & $7.82\pm 0.02$ &  $138^{+41}_{-26}$  & $3.0^{+0.5}_{-0.3} \times 10^{-5}$ \\
\enddata
\tablecomments{$^a$ The Cr He$\alpha$, Mn He$\alpha$ and Fe Ly$\alpha$ lines are narrow, and their widths cannot be constrained.}
\end{deluxetable}

We subsequently tried to fit the
spectrum with single thermal plasma 
models, including an absorbed 
CIE
model ($vvapec$) and an absorbed recombining plasma
model ($vvrnei$). 
The Tuebingen--Boulder interstellar medium (ISM) absorption model $tbabs$ is adopted to account for foreground absorption. The latest measurements of solar abundances ($aspl$ in XSPEC) are used \citep{asplund09}.
Neither of the single-temperature
models provides an acceptable fit to the Fe emission, despite
the fact that the $vvrnei$ model better describes the Fe Ly$\alpha$ emission (see the best-fit results in  Table~\ref{tab:fit}).
As shown in Figure~\ref{fig:4spec}, the Fe He$\alpha$ line (with a width of $45\pm1$~eV) is too wide 
to be explained with the two single-temperature models.
Using a velocity-broadened recombining model $bvvrnei$ improves the fit, but strong
residuals are still shown, giving a large $\chi_\nu^2=1.95$.
This prompted us to use  a two-temperature model, which is often suited for SNRs.

\subsection{Two-temperature model} \label{sec:doublecomp}

A two-temperature component is needed to fit the spectrum in
the 2--8 keV band, as pointed out in previous studies \citep{sakano04,park05, koyama07, one19}.
We found that an absorbed two-temperature plasma model $vvapec+vvapec$ with abundances tied together well
describes the spectrum in the 2--8~keV band and
provides a reduced $\chi^2$ ($\chi^2_\nu$) of 1.45 (dof=381).
This model suggests that 
the elements between the cool and hot components are well mixed. 
The best-fit results for the $vvapec+vvapec$ model with 90\% uncertainties are 
shown in Table~\ref{tab:fit}.

Here we explain why the abundances of the two components are tied.
First, we tested the unmixed case by freeing the abundances of IGEs in the hot component and fixing the abundances of 
IGEs in the cool component to the ambient value 
(1 or 2). This model results in a bad fit ($\chi^2_\nu\sim 2.8$),
because the spectral fit is sensitive to the Fe abundance in both components. A low Fe abundance in the cool
component causes large residuals in
the Fe He$\alpha$ and Ly$\alpha$ lines.
Hence, both the cold and hot components contain SN ejecta. 
Second,
the element abundances in the hot component 
cannot be independently constrained because of a strong degeneracy with the cold component. 
The best-fit abundances are mainly determined by the cold component, which dominates the photons below 6.7~keV (see Figure~\ref{fig:spec}).
Separating the abundances of the two components will invoke manual tuning of 
the hot component's abundances. This requires knowledge of how unmixed the 
ejecta is, which we do not know.
Finally, coupling the abundances of two ejecta components is based on the simplest and natural 
consideration: the ejecta is well mixed.
In this case, the cool component corresponds to the denser clumps, while the hot component is from the intercloud gas (see further discussion about the plasma in Appendix~\ref{sec:snr_properties}).  
Therefore, we 
tied the abundances between the cool and hot components and
allow the abundances of S, Ar, Ca, Cr, Mn, Fe, and Ni to vary.
This further supports the previous \XMMN\ and \Chandra\ studies, which used the same 
abundances in two-temperature components \citep{sakano04,park05}.

We also tested the underionized ($vvnei$) and recombining plasma models ($vvrnei$) for the cool and hot components.
We found that the hot component is in CIE, given the large ionization or recombining timescale ($>10^{13}~\cm^{-3} \s$) in the $vvnei$ and $vvrnei$ models. 
The cold component could also be fitted using an underionized 
model $vvnei$ with a
large ionization timescale of $\tau_c >0.7\E{12} \cm^{-3}\s$, implying a CIE condition \citep{smith10}.
The $vvnei+vvapec$ two-temperature model gives best-fit parameters and $\chi_\nu^2 (=1.43$) similar to the $vvapec+vvapec$ model. 
Therefore, it is equivalent to use the $vvapec+vvapec$ model as the best model.

Finally, we compared our results with earlier X-ray studies of \snr.
Our best-fit model, plasma temperature, and metal abundances of IME and Fe are consistent with the previous Chandra study by \cite{park05}, but we added the Cr, Mn, and Ni abundances and provided
better constraints for all parameters with deep \Chandra\ observations. 
Our best-fit temperatures are similar to that from the \XMMN\ study by \cite{sakano04}, which also suggests low abundances of S, Ar, and Ca ($Z/Z^\odot=1$--3).
We found larger differences between our \Chandra\ results  
and the \Suzaku\ results by \cite{one19},
which  requires two recombining plasma models with different IGE abundances.
This large discrepancy is likely caused by the different spectral extraction regions. 
The elliptical region shown in Figure~\ref{fig:x_c} has a size of $1'\times 1\farcm{5}$, while
the \Suzaku\ spectrum in \cite{one19} was selected in a $3\farcm{2}$-diamter region.
Due to the larger point spread function of \Suzaku\ \citep[$\sim 1'$ or worse,][]{one19},
the \Suzaku\ spectrum of \snr\ suffers more contamination from
background structures such as nonthermal filaments, bright point-like sources,
and other irrelevant emission near Sgr A*.

The previous \Chandra\ and \XMMN\ studies have a revealed a 
spatial variation of
physical parameters across \snr.
This paper focuses on the global properties of \snr, especially for the abundance ratios. 
In the appendix, we have extracted spectra from three regions
across \snr. Despite a variation of gas temperature, the abundance ratios of the IGEs are still consistent with the global spectrum.

\section{Discussion}

\subsection{Comparison with SN nucleosynthesis models }

The abundance pattern of \snr\ suggests that its 
progenitor SN produced mainly IGEs 
but little IMEs of S, Ar, and Ca.
This pattern is surprising and is remarkably distinct from 
other known SNRs.
The two main categories of SNRs are Type Ia SNRs
from C+O WDs and core-collapse SNRs from massive stars. 
Both SN categories produce a moderate 
amount of IMEs relative to IGEs,
and especially core-collapse SNe produce much more
IMEs.

\begin{figure*}
  \centering
\includegraphics[angle=0, width=0.49\textwidth]{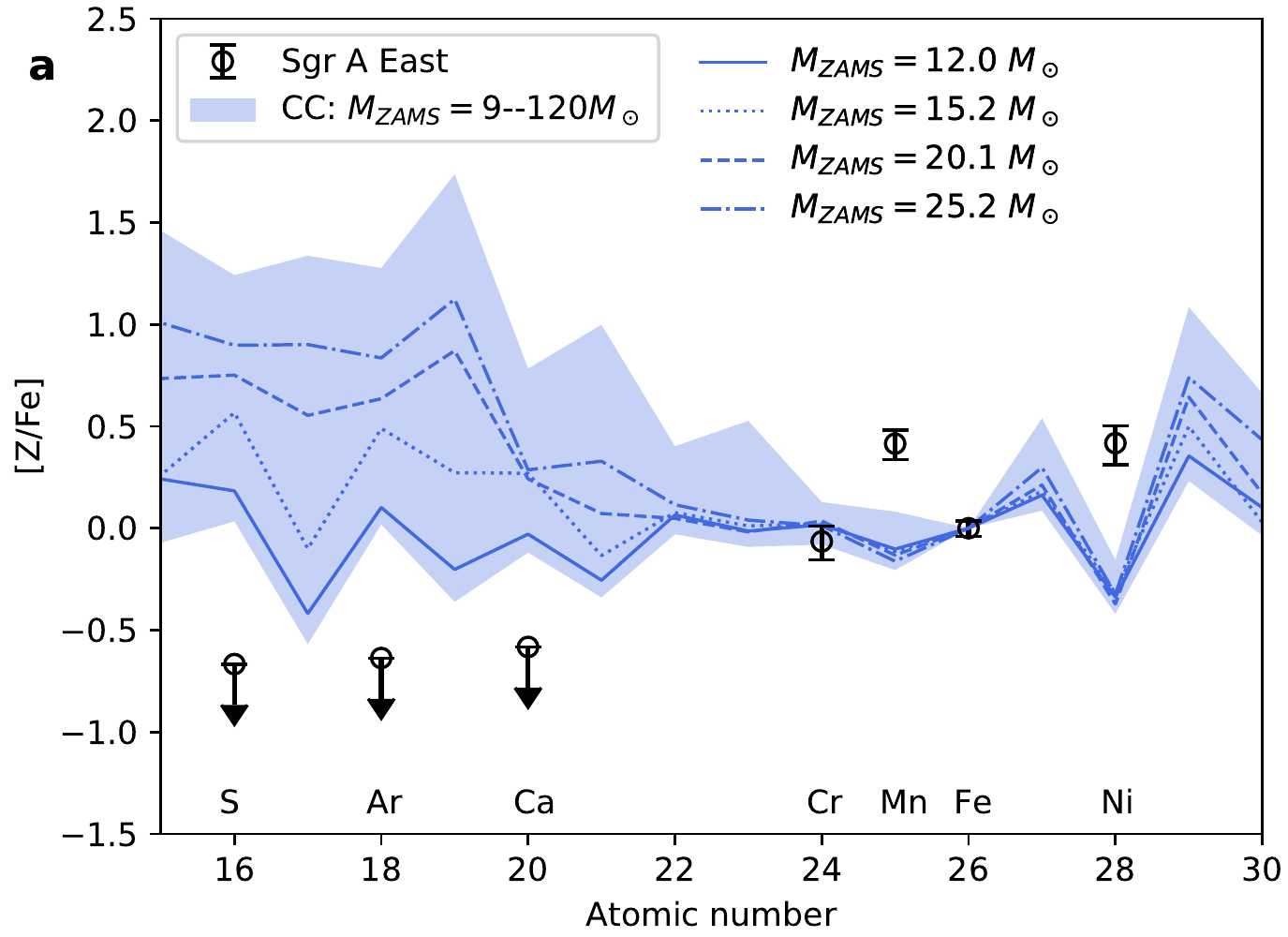}
\includegraphics[angle=0, width=0.49\textwidth]{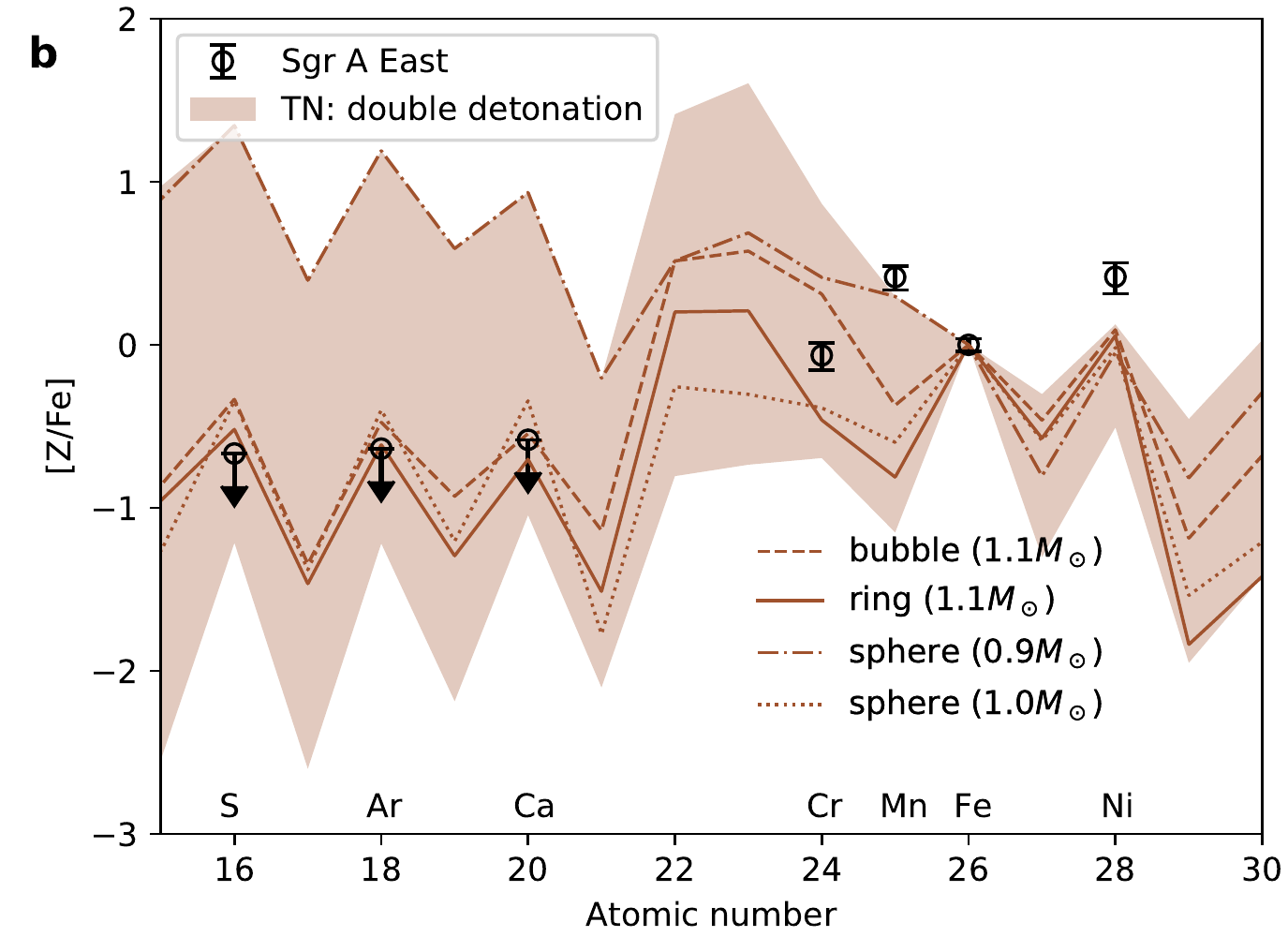}
\includegraphics[angle=0, width=0.49\textwidth]{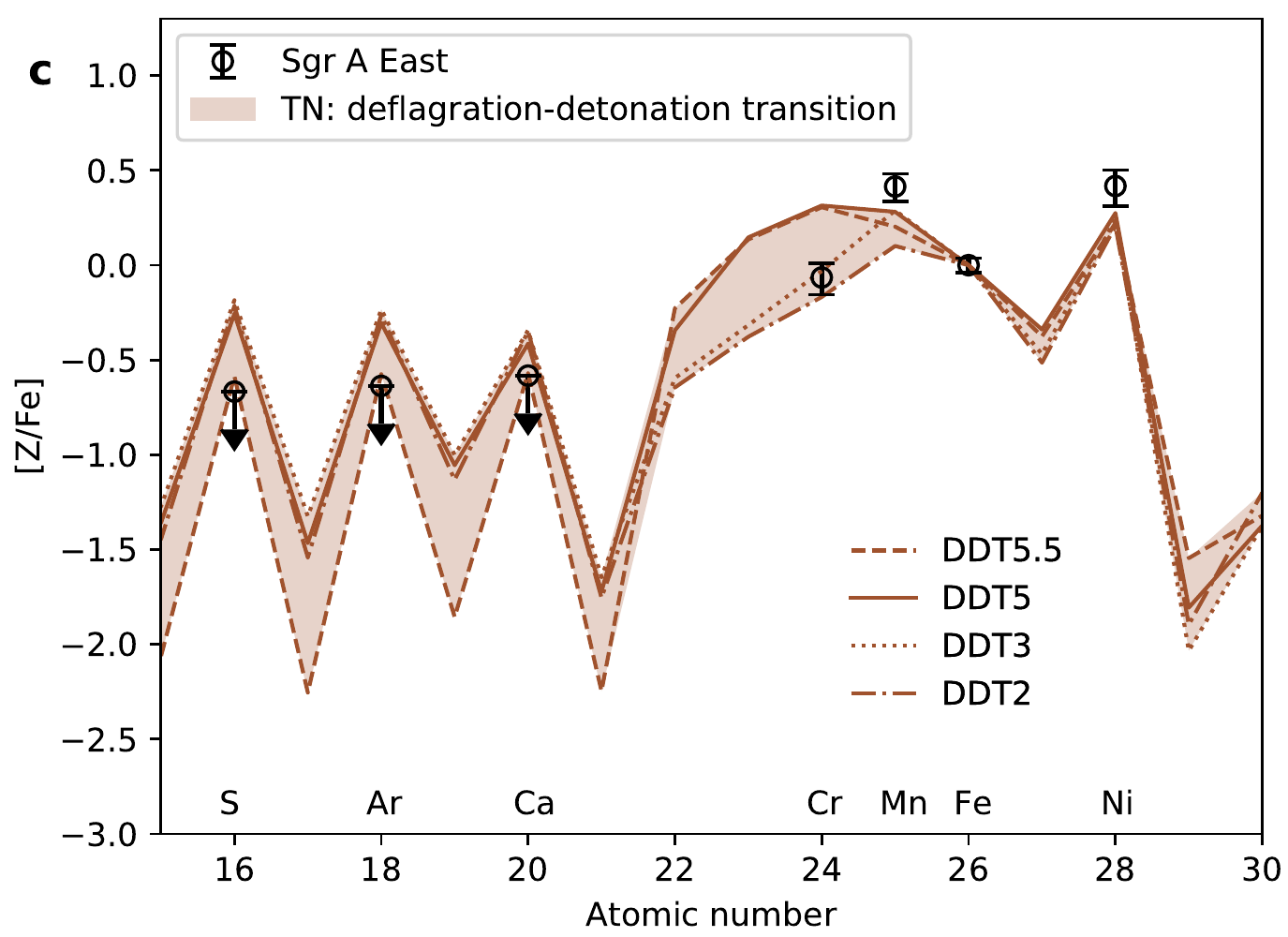}
\includegraphics[angle=0, width=0.49\textwidth]{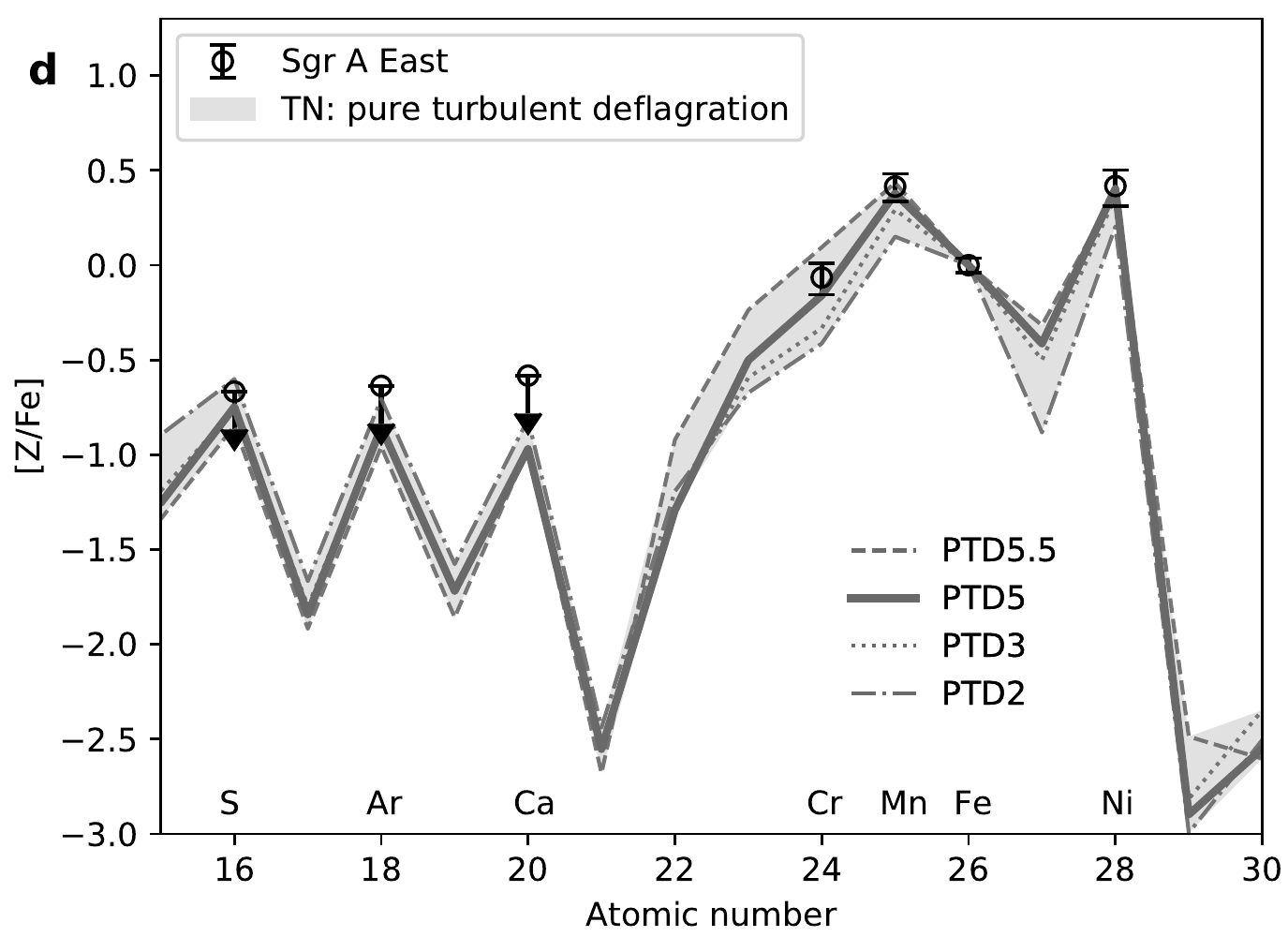}
\caption{
 A comparison between the observed logarithmic abundance ratios (relative to Fe)
and core-collapse (CC) and thermonuclear (TN) nucleosynthesis  models. The error bars give 90\% uncertainties.
(a) core-collapse SN models for stars with zero-age-main-sequence masses from 9 to 120 $M_\odot$ \citep{sukhbold16}.
(b) double detonation 
Type Ia models
for sub-Chandrasekhar-mass WDs (0.9--1.2~$\Msun$) with three initial He detonation configurations \citep{leung20subchand}. Three benchmark models and a low-mass WD model ($0.9~\Msun$ WD) are labeled with lines.
(c) deflagration-detonation transition (DDT) Type Ia SN models for
C+O WDs with central densities of 
$2\times 10^{9}$ (DDT2), $3\times 10^{9}$ (DDT3), 
$5\times 10^{9}$~g~cm$^{-3}$ (DDT5), 
and $5.5\times 10^{9}$~g~cm$^{-3}$ (DDT5.5), respectively \citep{leung2018Chand}.
(d)  pure turbulent deflagration (PTD) Type Iax SN models \citep{leung20Iax} for
WDs with central densities in the range of 2.0--$5.5\times 10^9$~g~cm$^{-3}$
(PTD2--5.5).
The thick solid line shows the best-fit Type Iax model with a central density of $5\times 10^9$~g~cm$^{-3}$.
}
\label{fig:progenitor}
\end{figure*}

In order to probe the explosion mechanism of \snr, we compare 
the logarithmic abundance ratios in
\snr\ with those predicted in several nucleosynthesis models
of SNe (see Figures \ref{fig:progenitor} and \ref{fig:progenitor_multispot}).
The logarithmic abundance ratio of element Z to Fe is defined as [Z/Fe]= $\log_{10}(Z_{\rm Z}/Z_{\rm Fe})-\log_{10}(Z_{\rm Z}/Z_{\rm Fe})^\odot$.
\snr\ has logarithmic abundance ratios [Cr/Fe] $=-0.065^{+0.075}_{-0.091}$,
[Mn/Fe] $=0.41^{+0.07}_{-0.08}$,
[Ni/Fe] $=0.42^{+0.08}_{-0.11}$,
[S/Fe] $<-0.60$, [Ar/Fe] $<-0.58$, and [Ca/Fe] $<-0.55$.
In this paper, we also use abundance ratios defined as
Z/Fe = $(Z_{\rm Z}/Z_{\rm Fe})/(Z_{\rm Z}/Z_{\rm Fe})^\odot$.

The contamination of the ambient gas to the SN ejecta is considered in
the comparison.
The clearly enhanced abundances of IGEs support that
the IGEs are dominated by the ejecta component.
Contrarily, the S, Ar, and Ca abundances are 
close to the average values in the Galactic center \citep{mezger96, rudolph06, davies09}, 
indicating small S/Ar/Ca yields from the SN and a strong dilution with the ISM.
Therefore, the [(Cr, Mn, Ni)/Fe] ratios represent the ejecta
values and can be directly compared with the models.
Since the S, Ar, and Ca abundances are close to the ambient value, 
the abundance ratios of IMEs to Fe in
Figures~\ref{fig:progenitor} and \ref{fig:progenitor_multispot} are given as the upper limits.

\subsubsection{Core-collapse models}
In Figure~\ref{fig:progenitor}a, we compared the observation with
core-collapse SN models by \cite{sukhbold16} \citep[see also][for independent and consistent yields]{nomoto13}.  It shows that the
core-collapse SN origin of
\snr\ can be ruled out as it would overproduce 
IMEs relative to Fe and result in a too-small 
[Mn/Fe] ($-0.22$ to $0.11$) compared to the observed value of $0.41^{+0.06}_{-0.07}$.

\subsubsection{Type Ia models}

The observed high ratio of Mn/Fe is an important indicator for the progenitor
of a thermonuclear SN, which provides conditions for the
neutronization by electron capture.
Such ahigh Mn/Fe eliminates a sub-Chandrasekhar-mass WD progenitor 
because the low-density matter cannot produce both a high Mn/Fe ratio and sufficient $^{56}$Ni for normal Type Ia SNe
\citep{seitenzahl13a, shen18,leung20subchand}.
In contrast, near-Chandrasekhar-mass C+O WDs have 
high-enough central densities for the formation of a larger amount of Mn. 

Figure~\ref{fig:progenitor}b shows a comparison between \snr\ and the sub-Chandrasekhar-mass WD models with WD masses of 0.9--$1.2~\Msun$ and 
a solar metallicity \citep{leung20subchand}.
In these double detonation (DD) models,  a C detonation is triggered by an initial He detonation with three different configurations.
The benchmark Type Ia models  are  labeled with solid, dashed, and  dotted lines. 
They represent typical sub-Chandrasekhar-mass models, reproducing $\sim 0.6~\Msun$ $^{56}$Ni masses and Type Ia SN explosion energies.
None of them explains the  observed high [Mn/Fe] and [Ni/Fe] ratios.

Among the 19 DD models,  
two low-mass (0.9~$\Msun$ WD) models result in oversolar [Mn/Fe] and [IMEs/Fe] ratios. 
They are not favorable models for normal Type Ia SNe as
they produce too-small $^{56}$Ni mass
($\le 0.14~\Msun$). 
Moreover, they fail to explain the low IME abundances observed in \snr.

We remind readers that \cite{leung20subchand}
has done a detailed survey for the parameter dependence of SNe Ia using
sub-Chandrasekhar-mass WDs.
The WD mass and metallicity (0--5$Z_\odot$) are the primary parameters that affect the global abundance pattern. To achieve the supersolar [Mn/Fe] ratio, for sub-Chandrasekhar-mass WD, a low-mass or high-metallicity model is necessary. 
However, such a model cannot explain the subsolar IMEs/Fe ratios observed in \snr, where the model shows at least 1 -- 2 dex above solar values.

We thus compared \snr\ with the near-Chandrasekhar-mass models for
Type Ia and Type Iax SNe.  In
these models, the turbulent deflagration wave
propagates initially at a slow subsonic speed from the WD center \citep{nomoto76,nomoto84}.
In the deflagration-detonation transition (DDT) models by \cite{leung2018Chand} adopted here \citep[see][for a review]{ropke07}, the transition from the
deflagration to the supersonic detonation is assumed to occur at a 
density as low as $\sim 10^7$ g cm$^{-3}$.  In the pure turbulent deflagration (PTD) models, it is
assumed that the transition from the deflagration to detonation does not occur.

For comparison with \snr, we applied the theoretical yields of the
two-dimensional DDT models \citep{leung2018Chand} and newly
calculated two-dimensional PTD models \citep{leung20Iax}.
These yields are given in Tables~\ref{tab:ddt_dd}--\ref{tab:ptd} in the appendix.
Our discussion focuses on these two recent models with centered flame structures, followed by a brief comparison with earlier models with other flame structures by \cite{seitenzahl13b} and \cite{fink14}, which assumed that the WDs are ignited with multiple flaming bubbles.

Figure~\ref{fig:progenitor}c shows the DDT models with central densities of $2$--$5.5\times 10^9$ g~cm$^{-3}$ 
\citep[models DDT2--5.5,][]{leung2018Chand}. Here
we added DDT2 and DDT5.5, which were not included in  \cite{leung2018Chand}. 
We noted the following from Figure~\ref{fig:progenitor}c: 
(1) DDT2, DDT3, DDT5, and DDT5.5 produce
[Mn/Fe] and [Ni/Fe] values that are a little smaller than the observation;
(2) DDT5 and DDT5.5 produce too-large [Cr/Fe] to be compatible with the
observation;
(3) DDT2 and DDT3 reproduce the observed [Cr/Fe] but overproduce IMEs relative to Fe.
Thus, no model can explain the observed abundance pattern.

\begin{figure*}
  \centering
\includegraphics[angle=0, width=0.49\textwidth]{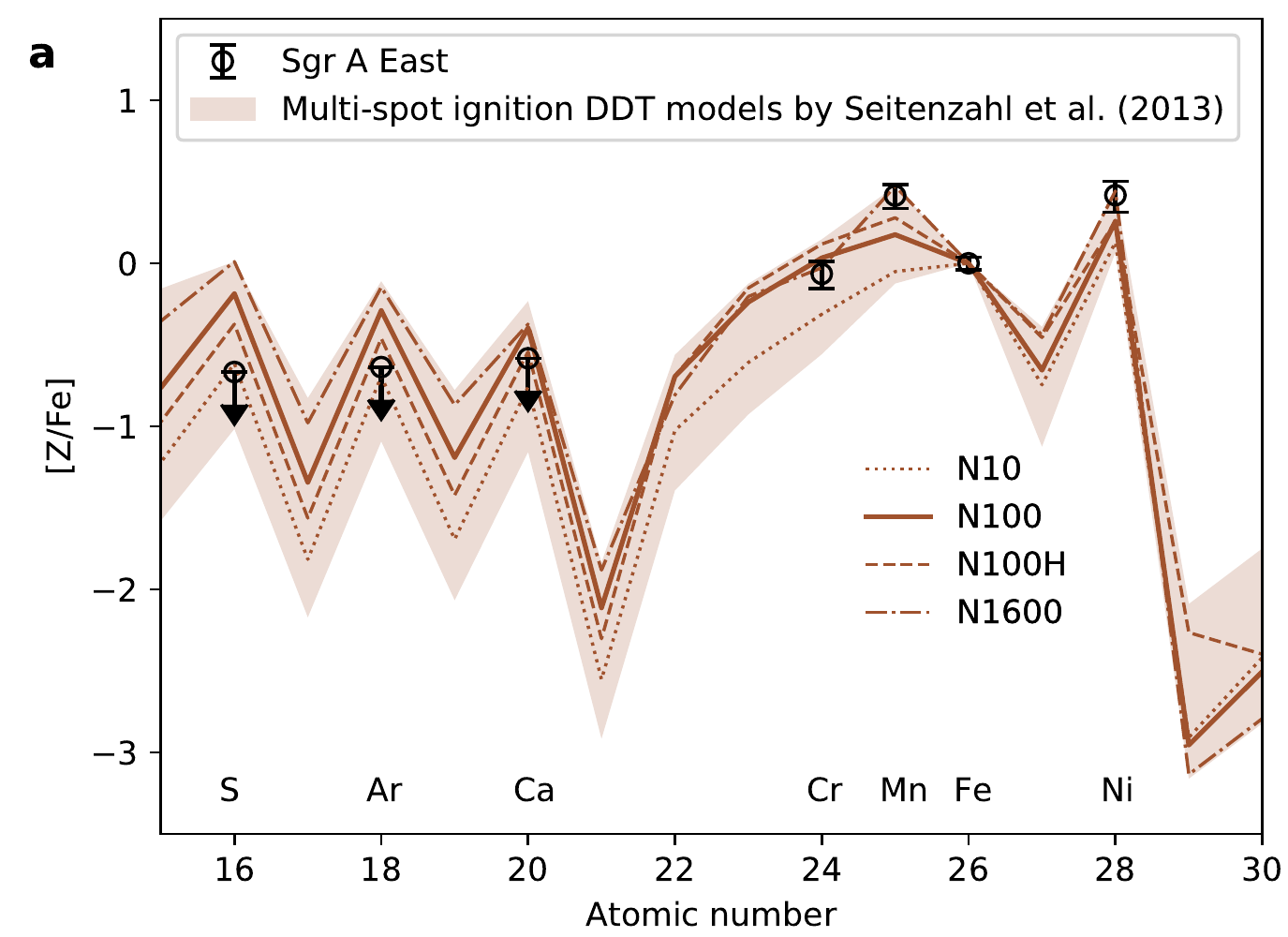}
\includegraphics[angle=0, width=0.49\textwidth]{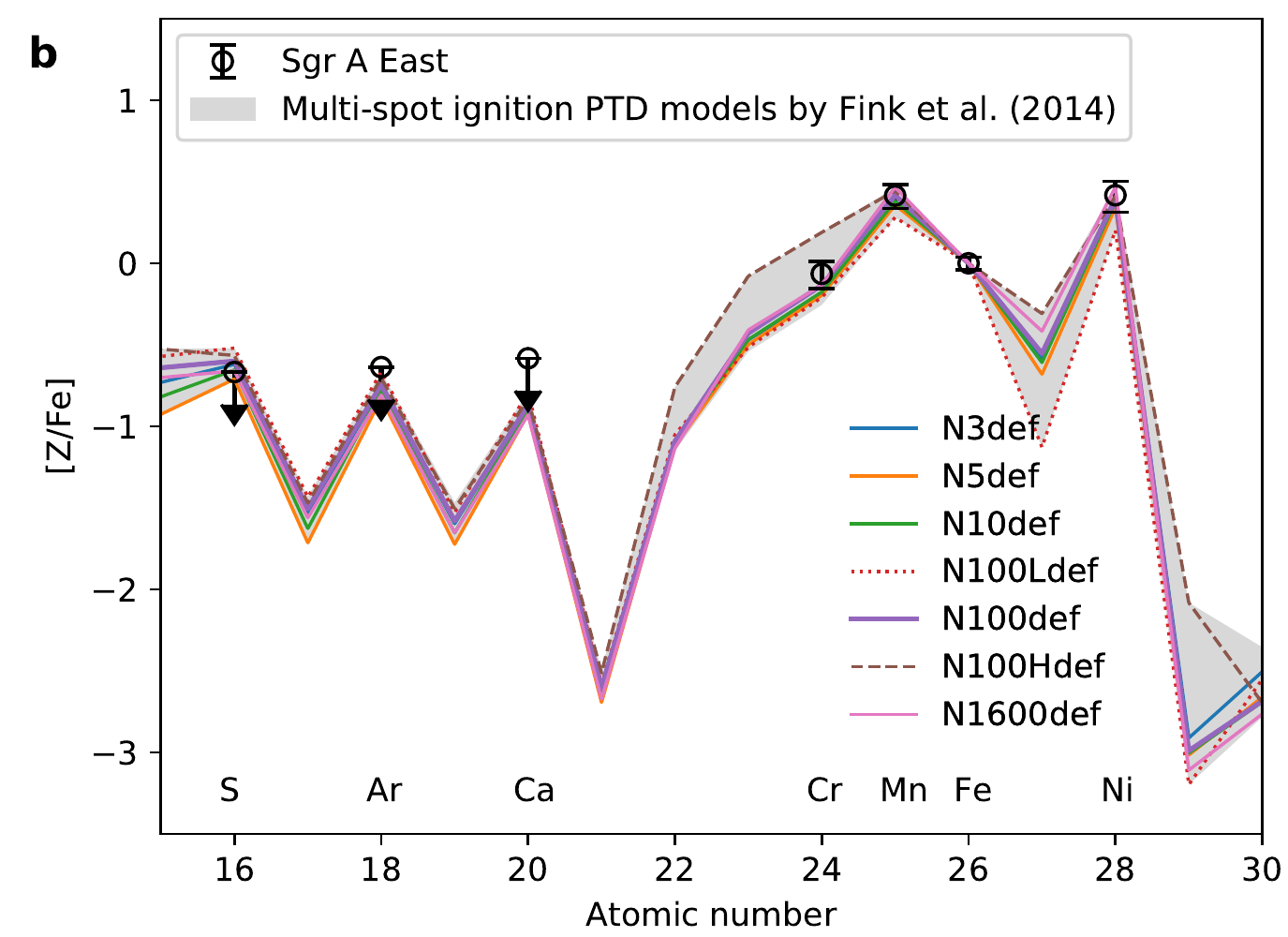}
\caption{
 Comparison between the observed logarithmic abundance ratios (relative to Fe)
and offcentered mutispot ignition DDT (a) and PTD models (b).
All available models have been considered in the shaded region, but 
a few exemplified models are shown with lines.
N10(def) and N100(def) mean 10 and 100 ignition bubbles, respectively.
In panel (b), the initial central density of the WD is $2.9\E{9}~\g\cm^{-3}$ for the N3--1600def models, $10^{9}~\g\cm^{-3}$ for the N100Ldef model and $5.5\E{9}~\g\cm^{-3}$ for the N100Hdef model.
}
\label{fig:progenitor_multispot}
\end{figure*}

We noted that the development and assumptions in SN nucleosynthesis models could affect the interpretation of the SNR origin.  
To investigate the dependence on hydrodynamics and nucleosynthesis models, 
we also compared with the three-dimensional DDT models from \cite{seitenzahl13b}, which assumed that the Chandrasekhar-mass WDs are ignited by multiple off-center flaming bubbles (up to 1600).
None of the models could explain both the low abundances of the IMEs and the large Mn/Fe ratio in \snr (see Figure~\ref{fig:progenitor_multispot}a). 
The N1600 model for strong deflagration with 1600 ignition bubbles is the only one reproducing the IGE/Fe ratios, but it largely overproduces IME/Fe ratios.

\subsubsection{Pure turbulent deflagration Type Iax models}

Lastly, in Figure~\ref{fig:progenitor}d, we examined the PTD models, 
which have been suggested to be the most likely mechanism  to explain 
the properties and large observational diversities in Type Iax 
SNe \citep[e.g.,][]{branch04,fink14,leung2018Chand}. 
Here we studied the dependence on the central density of the WD by
adopting the two-dimensional hydrodynamical models PTD2, PTD3, PTD5, and PTD5.5 for the central
densities of 2, 3, 5, and 5.5 $\times 10^9$ g~cm$^{-3}$, respectively \citep[Table 6 and][]{leung20Iax}.
According to these PTD models, the subsonic slow propagation
of the turbulent deflagration wave cannot sustain its burning and quenches, because 
the density at the burning front decreases with the expansion of the
WD.  Then the WD cannot be completely disrupted, and
a low-mass WD remnant is left behind the SN explosion.
The absence of supersonic detonation suppresses the formation of IMEs at
such low densities as $\sim 10^7$ g cm$^{-3}$.
Because of turbulent mixing associated with the deflagration, the
ejecta of PTD models contain IGEs that are synthesized in the high-density central region of the WD.  Compared with DDT models, the
ejecta of PTD models show much smaller IME/Fe ratios, which are
consistent with the observed upper limits to these ratios.

From Figure~\ref{fig:progenitor}d, we found that the ratios of [(Cr,
 Mn, Ni)/Fe] observed in \snr\ are well explained by the PTD5 and
PTD5.5 models, that is, the explosions of WDs with central
densities of 5 and 5.5 $\times 10^9$~g~cm$^{-3}$, respectively.
The best-fit model PTD5 (see Figure~\ref{fig:progenitor}d)
predicts that the explosion energy is $5.1\times 10^{50}$~erg and 
the $^{56}$Ni mass is $0.32 M_\odot$ \citep{leung20Iax}.
These predictions could explain a relatively bright Type Iax SN
\citep{foley13,stritzinger15}.

The central densities at the deflagration of the accreting WDs
basically depend on the accretion rate \citep{nomoto18}.
The densities of the PTD5 and PTD5.5 models ($5$ -- $5.5\times
10^9$~g~cm$^{-3}$) are somewhat higher than the typical central
densities ($2 - 3\times 10^9$~g~cm$^{-3}$) of normal Type Ia SN
models.  Such high densities at the deflagration
could be 
obtained by the thermonuclear runaway of a uniformly rotating 
WD ($5 - 6\times 10^9$~g~cm$^{-3}$) whose mass reaches $\sim 1.43~
M_{\odot}$ in the (so-called) spin up -- spin down scenario
\citep{benvenuto2015}.

The PTD5 model 
can bring an ejecta mass of $\sim 1.3~M_{\odot}$. This includes
a significant amount of cold fuel made of $^{12}$C and $^{16}$O. In fact, the ejecta mass is sensitive to how the initial flame is arranged. The PTD2--5.5 models assumed the flame starts at the center with angular perturbation. This flame structure refers to that presented in \cite{Reinecke1999,Reinecke2002a} for a robust mass ejection without invoking the detonation transition. The centered flame allows the flame to steadily burn the core for the steady expansion before the flame is quenched. The initial burning of the core provides the necessary thermodynamical conditions for electron capture and synthesis of neutron-rich isotopes, including $^{55}$Mn. 

Other flame structures, such as off-center flame bubbles, are presented in \cite{fink14}. The off-center flame encountered strong a buoyancy force, which drags the flame away from the center before it can sweep through the matter around the core. A drastically lower ejecta mass can result.

In Figure~\ref{fig:progenitor_multispot}b, the abundance ratios of
PTD models by \cite{fink14}  are compared
with \snr.  
These models assumed that the WDs 
are ignited by many
bubbles with number $N_k=1$ -- 1600. 
In the models N3def--N1600def (solid lines) with a WD central density
of $2.9\times 10^9$ g cm$^{-3}$, 
the predicted abundance patterns are marginally consistent with that in \snr. 

We then compare with N100Hdef, which
has a central density of $5.5 \times 10^9$ g cm$^{-3}$ at the ignition.  
It predicts a higher [Cr/Fe], with the Cr and Fe yields 
similar to the PTD5.5 model by \cite{leung20Iax}.
The comparison in Figure~\ref{fig:progenitor_multispot}b implies a central density of 3--$5 \times 10^9~\g~\cm^{-3}$ for the WD progenitor of \snr, marginally consistent
with the central density suggested by the PTD5--5.5 models (see Figure~\ref{fig:progenitor}d).
This agreement reinforces that
\snr\ has an abundance pattern typical for PTD of WDs, no matter how
the initial flame structures appear.

\cite{fink14} made comparisons of models with the observed light curves
and spectra of SNe Iax and found that the weaker and fainter explosion models with
$N_k=5$ and 10 are better than larger $N_k$ models for explaining SN~2002cx-like Type Iax SNe.
\cite{long14} also performed a three-dimensional simulation for multispot PTD WDs, but they suggested a different trend: few-$N_k$ models producing stronger and brighter explosions \citep[see][for an interpretation of the difference]{fink14}.

Figure~\ref{fig:progenitor_multispot}b implies that the abundance
pattern, especially for [Cr/Fe], is not sensitive to the propagation velocity of the
deflagration wave (related to $N_k$) but is sensitive to the central density.
It should be useful to construct high-central-density models, like
PTD5 and PTD5.5, but with a slower deflagration (small $N_k$) that leads to a wide
range of ejected mass (e.g., $^{56}$Ni) and explosion kinetic energy.  
Since Type Iax SNe show a large variation of light
curves and spectra, it would be interesting to find a range of
models that can explain the wide range of observed properties of
Type Iax SNe and also the chemical abundance pattern like in Sgr
A East.  The largest differences among the models are the metal
yields, but our X-ray study cannot constrain the metal masses (a lower
limit for Fe mass is $M_{\rm Fe}\sim 3\E{-2}~\Msun$; see
Appendix~\ref{sec:snr_properties}).

\subsubsection{Other remarks on the models and abundance ratios}

This paper mainly compares solar-metallicity nucleosynthesis models for consistency\footnote{The CC models for 9--12~$\Msun$ stars assume zero metallicity \citep[see details in][]{sukhbold16}. For the metallicity dependence of the CC supernova yields, see \cite{nomoto13}.}.
The full DD and DDT models in \cite{leung20subchand,leung2018Chand} have covered a wide range 
of  progenitor metallicity ($Z=0$--5$Z_\odot$). 
These non-solar-metallicity models with WD central densities of 1--$5\E{9}~\g\cm^{-3}$ still fail
to explain the abundance pattern in \snr.
Moreover,
the observed low IME abundances ($Z_{\rm S}/Z_{\rm S}^\odot=1.4\pm 0.2$) imply that high-metallicity models 
are not necessary for \snr.
For the Type Ia and Iax models, the dependence on parameters, such as the central density, metallicity, and input physics, has been extensively studied \citep[see][with a discussion about a few other groups' work]{leung20subchand, leung2018Chand, leung20Iax}.
Appendix~\ref{sec:simulation} provides further details about the numerical simulations of these models.
Our paper has compared four explosion mechanisms, but the future development of SN models may bring a wider view. 

\begin{figure}
\epsscale{1.15}
\plotone{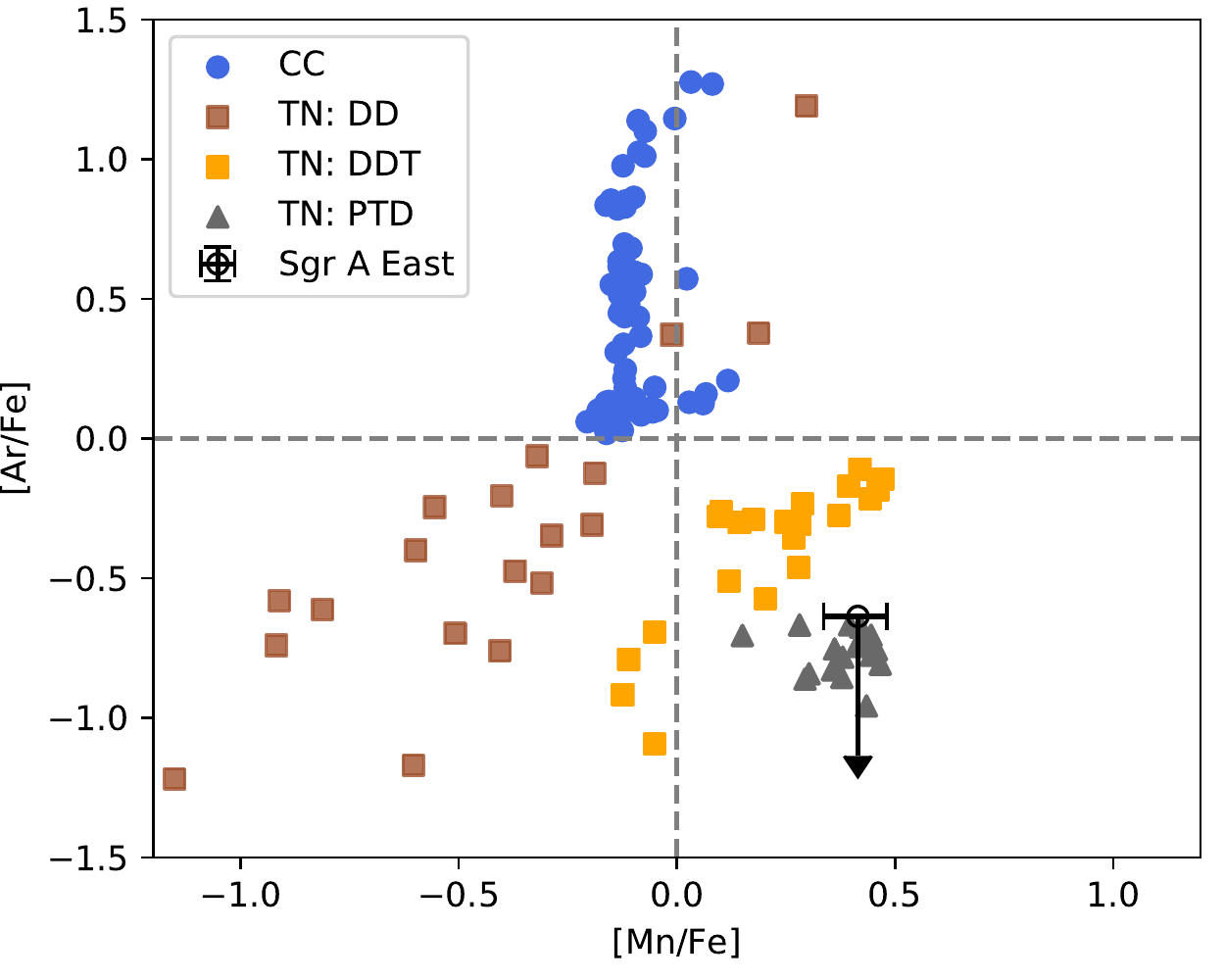}
\caption{
[Ar/Fe]--[Mn/Fe] diagram for four groups of explosion mechanisms plotted
in Figures~\ref{fig:progenitor} and \ref{fig:progenitor_multispot}.
The observed ratios in \snr\ are overplotted for comparison.
The dashed lines denote the solar values for [Mn/Fe] and [Ar/Fe].
}
\label{fig:ratio_mn_ar}
\end{figure}

The four explosion mechanisms used in our study can be distinguished 
in the  [Ar/Fe]--[Mn/Fe] diagram (Figure~\ref{fig:ratio_mn_ar}).
CC SNRs are mainly distributed in the second quadrant, but
some may appear in the first quadrant.
The DD models for sub-Chandrasekhar-mass WDs do not
appear in the fourth quadrant.
The PTD points are concentrated at the lower right region
in the fourth quadrant, well separated from the DDT points.  \snr's position in the [Ar/Fe]--[Mn/Fe] diagram favors a PTD explosion mechanism. 
While the diagram gives useful clues,
we still suggest using a large group of elements 
to distinguish explosion mechanisms.

\subsection{Type Iax origin}

Our study suggests that \snr's abundance pattern is consistent 
with a PTD explosion of a near-Chandrasekhar-mass WD,
a leading mechanism of producing Type Iax SNe.
This suggests \snr\ could be the first identified Type Iax SNR. 
Type Iax SNe occur at a rate of $31^{+17}_{-13}$  for every 100 Type Ia SNe
in a given volume \citep{foley13}.
Given the large occurrence rate, we should expect to find a few or more Galactic SNRs with Type Iax origin.
However, only \snr\ has been identified so far, likely due to the difficulty of
IGE measurements in old SNRs, especially for the faint lines of Mn and Cr.
In our Galaxy, we have known of three confirmed Type Ia SNRs (Tycho, Kepler, and SN~1006) 
and two candidates (RCW~86 and G1.9+0.3) younger than 2~kyr. 
If \snr\ is younger than 2~kyr \citep{rockefeller05, fryer06}, the Type Iax SNRs are found at a rate of 1 for 3--5 Type Ia SNRs in the past 2~kyr, which
is consistent with the measured Type Iax SN ratio
(see a detailed discussion about the dispute of the SNR age
in Appendix~\ref{sec:snr_properties}).

Despite the small statistics, one may wonder why the first Type Iax SNR 
is found near the Galactic center rather than other places  in our Galaxy.
The identification has been made possible by the deep {\it Chandra} observations
toward the Galactic center.
Type Iax SNe are preferentially found in a young
environment and arise from more massive WDs with short evolution time \citep{lyman13}.
It is suggested that the population of WDs in cataclysmic variables in the Galactic center have a mean 
mass of 1.2~$M_\odot$ (assuming non-magnetic WDs), 
significantly heavier than the WDs in local and Galactic-bulge cataclysmic variables \citep[$\sim 0.8~M_\odot$,][]{xu19}.
Although many Type Iax SNe were found in the ``outskirts" of late-type galaxies  \citep{jha19}, 
our study suggests that this peculiar class can also occur in the center of galaxies.

According to the PTD models, a bound remnant should be 
left in the explosion. It is of interest to
search for the WD survivor near \snr, but the high
absorption toward the Galactic center brings difficulties.
Recently, a few WDs with peculiar kinematics and spectroscopic properties have been identified in our Galaxy \citep{vennes17,raddi19}:
they are less massive than typical WDs (only 0.14--$0.28~\Msun$), are inflated (with radii of 0.08--$0.16R_\odot$), and show chemical similarities 
with the predicted survivors of Type Iax or thermonuclear electron-capture SNe. 
The studies on these inflated WDs provide another 
important angle for investigating peculiar thermonuclear SN explosions.

A few observations were used to indirectly argue the core-collapse origin of \snr. 
\snr\ was considered to be associated with the ``cannonball'' neutron star outside the
SNR boundary \citep{park05}, with a transverse velocity of  500~km s$^{-1}$ \citep{nynka13}.
This requires that the \snr\ be an old SNR with age $\sim 10^{4}$~yr, which is inconsistent with the hydrodynamic simulations of the SNR evolution \citep[$\lesssim 2$~kyr,][]{rockefeller05,fryer06}.
Our X-ray analysis does not give any evidence for a large age of the 
SNR. The large ionization timescale of 
\snr's plasma does not necessarily support a large SNR age
\citep[see Appendix~\ref{sec:snr_properties}, and see][for a few possible explanations of the high ionization state of the plasma]{one19}.
\cite{yalinewich17} found that \snr\ in the Galactic center evolves faster than other Galactic SNRs, and it cannot be an old SNR associated with the ``cannonball'' neutron star.
Furthermore, the abundance pattern of \snr\ suggests a WD survivor but 
excludes a neutron star survivor.
Therefore, the association of the ``cannonball'' has not been established 
to imply a core-collapse origin for \snr.

Several earlier X-ray studies considered a core-collapse origin for \snr\ \citep{maeda02,park05}, but these studies did not present a comparison between all of the constrained metal abundances and different SN nucleosynthesis models. 
As shown in Figure~\ref{fig:progenitor}, the core-collapse models 
have severe problems in \snr,
as they produce too much IMEs and too little Mn and Ni to Fe.
\cite{sakano04} pointed out that Type Ia may explain why Fe is more abundant than IMEs, although they did not exclude a Type II. 
\cite{park05} constrained the maximum Fe mass of $0.27~\Msun$ and argues that a low Fe mass does not favor a Type Ia. 
This Fe mass does not rule out a Type Iax origin, as Type Iax SNe show a range of $^{56}$Ni masses \citep[0.003 -- $0.3~\Msun$,][]{mccully14,stritzinger15}.
Knowledge of the Type Iax SN group has been gradually accumulated in the past decade \citep[since the discovery by][]{li03}. The Type Iax possibility was thus not examined for \snr\ through X-ray spectroscopy.
Moreover, we note that the measured Fe mass in \snr\ has a large uncertainty, depending on
the assumed geometry of the X-ray-emitting gas
(see Appendix~\ref{sec:snr_properties} and Section~\ref{sec:3c397_w49b}).

Infrared observations toward \snr\ have revealed $0.02~M_\odot$ of warm dust, which
was attributed to core-collapse SN dust surviving the passage of the reverse shock \citep{lau15}.
This was based on the opinion that core-collapse SNe are dust factories,  and there is no observational evidence for Type Ia SNe also producing a lot of dust \citep{sarangi18}.
The formation of SN dust depends not only on the composition of the ejecta but also on the ejecta velocity and temperature
\citep{nozawa03}.
The low shock velocities in Type Iax SNe create better conditions for dust condensation and formation of large dust grains  \citep[see][for Type Ia SNe]{nozawa11}.
According to our best-fit model PTD5,  0.31~$M_\odot$ of unburned C  
can be ejected in the Type Iax SN (see Table~\ref{tab:ptd} in the appendix).
The C mass is over two orders of magnitude larger than that produced in
Type Ia DDT models \citep[see][and Appendix Table~\ref{tab:ddt_dd}]{leung2018Chand},
and even larger than
the amount generated by a core-collapse SN from
the $20.1~M_\odot$ star \citep[see][]{sukhbold16}.
In the weak-explosion PTD models by \cite{fink14} with off-center flame structures, most of the unburned C is left in the
bound remnant, but the ejected C is still around one order of magnitude larger than
that in DDT Type Ia SNe. 
Our PTD models also predict a large amount of O and Fe elements for
bright Type Iax SNe.
Without dedicated studies for Type Iax SN dust, we do not
know how much ejecta could be condensed to dust grains and the composition of the dust.

The existence of dust in \snr\ does not exclude the thermonuclear origin of \snr.
Instead, this indicates that it is necessary to explore if Type Iax SNe from PTD explosions are potential factories of dust grains, because neither 
core-collapse nor Type Ia models explain the X-ray properties of \snr.
In our PTD5 model, if $\sim 6\%$ of the C (or other dust grains) condensed in the dust phase and survived the shocks, the Type Iax origin can explain 
the warm dust found in \snr.
A mid-infrared excess has been shown in the late-stage spectrum of Type Iax SN~2014dt and was interpreted as the newly formed dust from the SN ejecta or from the circumstellar medium
\citep{fox16}.
However, \cite{foley16} suggested that  
mid-infrared excess came from
a bound remnant with a super-Eddington wind.
Therefore, whether Type Iax SNe are dust
factories is still a question needing observational and theoretical tests.
Late-stage infrared monitoring observations of more Type Iax SNe are needed, especially for those brighter ones with larger mass ejection.
Another remark is that Type Iax SNe reveal a large diversity in their properties such as ejecta masses and velocities.
This means that we should not expect a uniform dust yield for this group.
Future {\it James Webb Space Telescope} ({\it JWST}) observations are expected to shed light on the composition of dust 
grains in \snr\ and test if Type Iax SNe are also dust producers.

\subsection{Comparison with 3C~397 and W49B} \label{sec:3c397_w49b}

Thermonuclear SNe are IGE factories. 
The nucleosynthesis models have predicted different metal outputs for various Type Ia and Type Iax SNe.
The measurement of IGEs in galaxies or the ISM is crucial for inferring
the diversity or population of the SNe \citep[e.g.,][]{seitenzahl13a}.
In this subsection, we compared the metal pattern of \snr\ with Type Ia SNRs, to provide observational evidence that they are different in metal outputs and that the thermonuclear SNRs are more than a uniform group. 

3C~397 and W49B are middle-aged, Fe-rich SNRs recently claimed to have Chandrasekhar-mass WD progenitors and DDT Type Ia explosions \citep{yamaguchi15, zhou18a}.
Both remnants were previously interpreted as core-collapse SNRs,
whose morphologies are strongly shaped by an interaction of dense atomic/molecular clouds
\citep[e.g.,][]{safi-harb05, keohane07,lopez13a}.
These properties are comparable to that of \snr. 
Another shared property among them is the centrally filled X-ray emission and shell-like
radio morphology, which categorize them as mixed-morphology SNRs \citep{lazendic06,vink12,zhang15}.
This is distinct from other young Type Ia SNRs with a shell-like X-ray morphology.
In the young Type Ia SNRs, such as Tycho, Kepler, and SN~1006, the inner portion of the 
ejecta has not been reheated by the reverse shock and thus is invisible in the X-ray band. This prevents us from making a good comparison of the chemical abundances between \snr\ and young Type Ia SNRs.

\begin{figure}
\epsscale{1.15}
\plotone{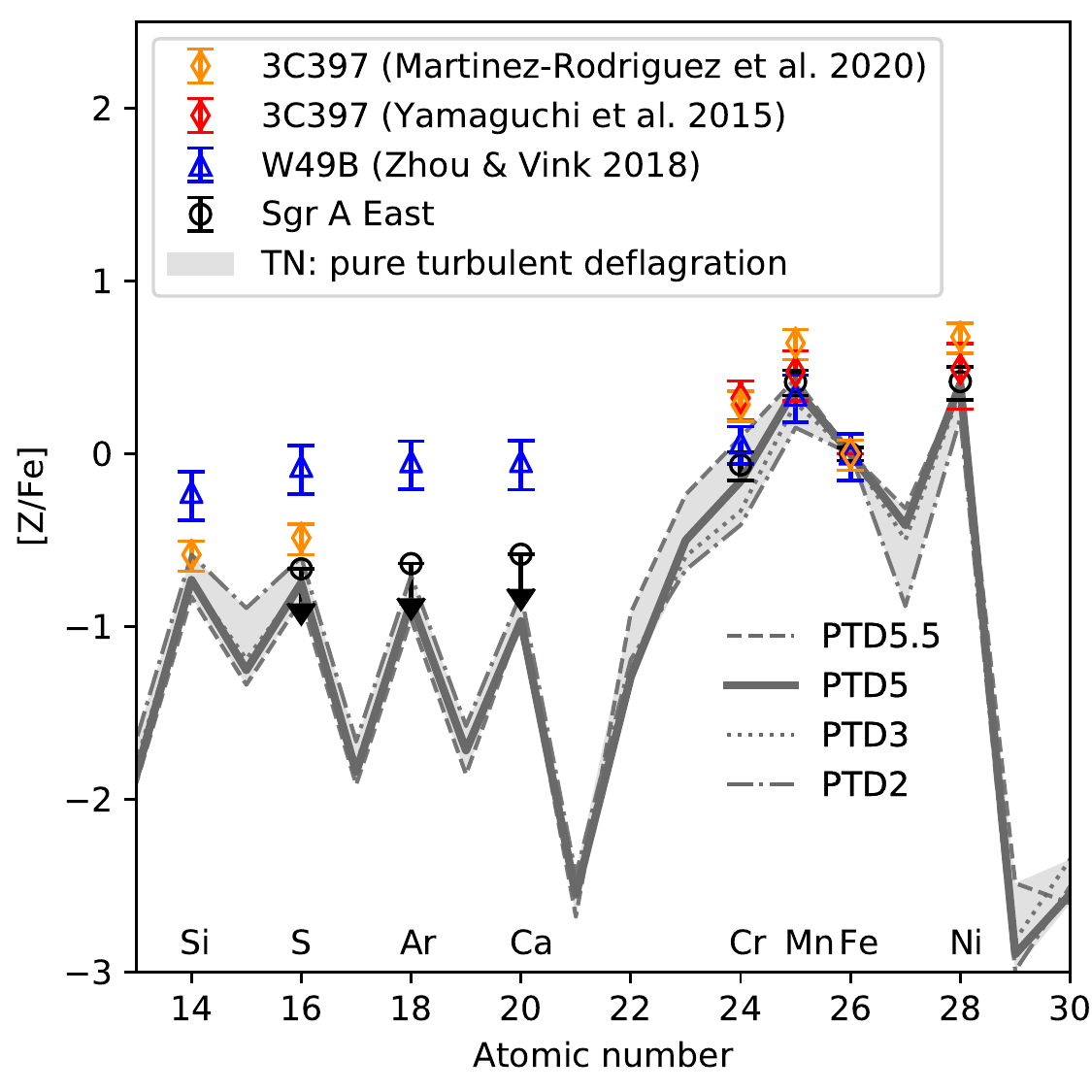}
\caption{
Comparison between the observed logarithmic abundance ratios (relative to Fe) for 3C~397, W49B, and \snr\ 
and the PTD models by \cite{leung20Iax}. 
We adopted a 20\% systematic error for the ratios from \cite{martinez-rodriquez20}, which have very small statistical errors.
}
\label{fig:ptd_3c397_w49b}
\end{figure}

Figure~\ref{fig:ptd_3c397_w49b} compares the logarithmic abundance ratios between \snr\ and 3C~397 and W49B over 
the PTD models.
It shows that \snr\ does not have the same abundance pattern as 3C~397 and W49B. 

\snr\ has a significantly lower [Cr/Fe] than 3C~397. 
The [Cr/Fe] value in 3C~397 is $0.32^{+0.10}_{-0.13}$ (1$\sigma$ error)  according 
to \cite{yamaguchi15} using 69 ks of \Suzaku\ data.
The value is refined to  $0.28^{+0.05}_{-0.09}$ (90\% error) 
in a recent study by \cite{martinez-rodriquez20} using 172~ks of \Suzaku\ data.
In contrast, \snr\ has a sub- or near-solar [Cr/Fe]$=-0.065^{+0.075}_{-0.091}$.
The [Mn/Fe] and [Ni/Fe] ratios in \snr\ are similar to the 3C~397 values determined by \cite{yamaguchi15}, but are smaller than that obtained by \cite{martinez-rodriquez20}.

Unlike \snr, 3C~397 has revealed high IME abundances. 
Previous studies of 3C~397 gave mass ratios $M_{\rm Ca}/M_{\rm S}=0.213^{+0.021}_{-0.034}$ and $M_{\rm Ar}/M_{\rm S}=0.214^{+0.030}_{-0.026}$, respectively \citep[90\% confidence range;][]{safi-harb05,martinez-rodriquez17}.
The recently updated abundance ratios 
gave $M_{\rm Ca}/M_{\rm S}=0.56^{+0.10}_{-0.15}$ and $M_{\rm Ar}/M_{\rm S}=0.40^{+0.05}_{-0.07}$ 
\cite[90\% confidence range,][]{martinez-rodriquez20}.
These new ratios are significantly larger than that predicted by
our PTD models ($M_{\rm Ca}/M_{\rm S}=0.12$--0.16 and $M_{\rm Ar}/M_{\rm S}=0.17$--0.18), while the old ratios are only slightly larger than the PTD results.
Due to the large discrepancy of the ratios between studies, Figure~\ref{fig:ptd_3c397_w49b} does not show [Ca/Fe] and [Ar/Fe] ratios for 3C~397.
Moreover, the [S/Fe] ratio in 3C~397 (considering a 20\% systematic error) is slightly larger than that of \snr.

The progenitor of 3C 397 has been suggested to be a high-metallicity ($Z\sim 5 Z_\odot$), high-density Chandrasekhar-mass WD, which
reproduces the large Mn/Fe and Ni/Fe ratios 
through a DDT explosion
\citep{yamaguchi15,leung2018Chand}.
Nevertheless, \cite{dave17} proposed another solution.
In addition to the high-Z, high-$\rho$ DDT models,
they found that a PTD explosion of a low central density ($\sim 2\E{9} \g\cm^{-3}$) WD could also reproduce the IGE masses in 3C~397 \citep[see also][]{leung2018Chand}.
Our PTD models with central densities of 2.0--$5.5\E{9}~\g \cm^{-3}$ cannot describe the high [Cr/Fe], [Ca/S], and [Ar/S] in 3C~397.
The high-density PTD model N100Hdef in \cite{fink14} gives [Cr/Fe]$=0.19$, and thus may marginally explain the high [Cr/Fe] ratio,
but it is still difficult to describe the oversolar [Ca/S] and [Ar/S] ratios.

W49B has been proposed as a remnant from a DDT explosion,
given its abundance pattern and the large observed IGE masses 
according to spatially
resolved spectroscopic analysis \citep{zhou18a,siegel20}.
Figure~\ref{fig:ptd_3c397_w49b} shows
that W49B produces too-large IME/Fe ratios compared to \snr, although the [Cr/Fe] and [Mn/Fe] values are similar to that of \snr.

Some disputes remain on the explosion mechanisms of W49B \citep{sun20,sato20}.
One major dispute was from the inconsistent metal masses obtained in different studies, although
the abundance ratios are more or less consistent \citep{sun20, siegel20}.
When considering low IGE masses,
the abundance pattern of W49B (except the high Mn abundance) might also be explained with a core-collapse SNR.
We stress that \snr's IME/Fe, Mn/Fe, and Ni/Fe ratios eliminate core-collapse origins (see Figure~\ref{fig:progenitor}).
In the Appendix (section~\ref{sec:snr_properties}), we elaborate that the metal masses for \snr\ have large uncertainties (over one order of magnitude) and should not be used to distinguish the explosion mechanisms.
The main reason is that the derived metal masses are sensitive to the assumed three-dimensional gas geometry and 
ejecta--ISM mixing, which we do not clearly know.
This issue also applies to W49B's metal masses when
analyzing the global spectrum.
In the ejecta--ISM mixed case, the metal masses $M_{\rm X}$ were derived mainly using two parameters: the fitted emission measure $EM$ and the filling factor of the gas in the volume $f$ ($M_{\rm X}\propto \sqrt{EM f} $).
Consequently, an accurate mass calculation needs a good understanding of $f$.
\cite{sun20} fitted the global spectrum using a three-component model and obtained a very small ejecta filling
factor $f_{\rm ej}=6.4\%$ and
a low Fe mass ($6.0\pm 0.6\E{-2}~\Msun$), too small
for a Type Ia SN.
In contrast, \cite{zhou18a} show that the gas in W49B is highly inhomogeneous and obtained a spatially varied $f_{\rm ej}$ with a mean value of $\sim 40\%$. They obtained significantly larger IGE masses 
($M({\rm Fe})=0.3\pm 0.1~\Msun$).
Moreover, the existence of pure ejecta would significantly enhance the metal masses with the same $EM$ and cause extra uncertainties \citep{park05, greco20}.
Considering the difficulty of obtaining ccurate measurements of the metal masses using a global spectrum and current low-energy-resolution CCD
spectra \citep{greco20}, a good way
is to lean on the abundance/mass ratios, which are less
affected by the geometry and pure-ejecta assumptions.

The comparisons in Figures~\ref{fig:progenitor}-- \ref{fig:ptd_3c397_w49b} demonstrate the importance of using a large set of element ratios to constrain the SN models.
Using only three IGEs (e.g., Mn/Fe or Ni/Fe ratios alone) is insufficient to distinguish PTD and DDT origins in some cases.
\snr\ has a Chandrasekhar-mass WD progenitor, but with a different explosion mechanism from 3C~397 and W49B.
The differences among \snr, 3C~397, and W49B provide observational evidence that there is a diversity of thermonuclear SNRs from Chandrasekhar-mass WDs, and they shows that this diversity could be probed by X-ray spectroscopy.

\section{Conclusion}
We have performed an X-ray spectroscopic study of SNR \snr\ using
3~Ms of Chandra data.
The metal pattern of \snr\ can be well explained with a pure turbulent deflagration (PTD)
explosion of a Chandrasekhar-mass WD, a leading mechanism for producing Type Iax SNe.
\snr\ increases the diversity of the known thermonuclear SNRs and
provides a valuable target for the study of Type Iax SNe.

We are aware that our interpretation of \snr's progenitor highly depends on the existing SN nucleosynthesis models. Model development
is important for bringing broader possibilities for a comparison with observations, although
the models used in our paper have already covered a wide parameter space.
On the other hand, a better understanding of the SN ejecta masses and
spatial distribution also help to constrain SN explosion mechanisms.
Next-generation X-ray spectrometers with a high spectral resolution will provide crucial insight into the ejecta composition and masses in not only \snr\ but also other SNRs.

Our main results are summarized as follows:

\begin{enumerate}
    \item The \Chandra\ ACIS-S spectrum of \snr\ shows clear emission lines of the S, Ar, Ca, Cr, Mn, Fe, and Ni. The strong Fe He$\alpha$ and Ly$\alpha$ lines have been resolved.
   The abundances relative to the solar values and the 90\% uncertainties of S, Ar, Ca, Cr, Mn, Fe, and Ni elements are
    $1.4\pm 0.2$, $1.5\pm 0.2$, $1.7\pm 0.1$, 
    $5.6\pm 1.0$, $16.9\pm 2.6$, $6.5\pm 0.4$, and $17.0\pm 3.5$, respectively. 
    
    \item  A two-temperature plasma model is needed to fit the 2--8 keV spectrum. We found that the best-fit model is an absorbed two-temperature (near-)CIE model with the cool and hot component abundances tied together, consistent with that suggested in
    previous \Chandra\ and \XMMN\ studies \citep{park05,sakano04}.
    This model suggests that the S--Ni elements between the cool and hot components are well mixed.
    The two components have temperatures of $\sim 1.2$~keV and $\sim 4.3$~keV, respectively. The foreground absorption is $\sim 2.1\E{23}~\cm^{-2}$.

    \item 
    \snr\ shows a low ratio of IMEs to Fe and large Mn/Fe and Ni/Fe ratios.
    This abundance pattern does not accord with core-collapse SN models \citep{sukhbold16} or normal Type Ia SN models for sub-Chandrasekhar WDs (due to the high Mn/Fe ratio) or DDT of Chandrasekhar-mass WDs with DDT explosions \citep[models from][]{leung2018Chand, seitenzahl13b}.
    
    \item The metal composition unveils that \snr\ originated from a PTD explosion of a Chandrasekhar-mass WD \citep{leung20Iax,fink14}, a popular mechanism for Type Iax SNe.
    \snr\ is thus likely the first identified Type Iax SNR and provides the nearest target for studying this peculiar class of SNe.
    
    \item The existence of a significant amount of dust in Sgr A East \citep{lau15}, together with the low ejecta velocity and large C (also O and Fe) production predicted in the best-fit PTD5 models, implies that Type Iax SNe could be potential dust factories. This speculation needs future tests from modelings and observations.

    \item \snr\ shows an abundance pattern distinguished from  
    3C397 and W49B, which were claimed to result from DDT Chandrasekhar-mass WD explosions (with disputes).
    The X-ray spectroscopy of SNRs provides observational evidence that there are diverse explosion channels for Chandrasekhar-mass WDs.
\end{enumerate}

\acknowledgments
We are grateful to Mark Morris, the anonymous referee, and Boris Gaensicke for constructive comments.
This research used the observations taken from the {\it Chandra} X-ray Observatory. 
The analysis has made use of software package CIAO provided by the {\it Chandra} X-ray Center and spectral fitting package XSPEC.
P.Z. acknowledges the support from the support from the Nederlandse Onderzoekschool Voor Astronomie (NOVA), the NWO Veni Fellowship, grant No. 639.041.647 and NSFC grants 11503008 and 11590781. 
S.C.L and K.N. acknowledge support by the World Premier International Research Center Initiative (WPI Initiative), MEXT, Japan. 
S.C.L. also acknowledges support by grants HST-AR-15021.001-A and 80NSSC18K101.
Z.L. acknowledges support by the National Key Research and Development Program of China (2017YFA0402703) and NSFC grant 11873028.
K.N. acknowledges support by JSPS KAKENHI grant Nos. JP17K05382 and JP20K04024.
Y.C.\ acknowledges support by NSFC grants 11773014, 11633007, and 11851305.

\software{
ATOMDB \citep{smith01,foster12},
CIAO \citep[vers. 4.10,][]{fruscione06}, \footnote{https://cxc.cfa.harvard.edu/ciao/index.html}
DS9,\footnote{http://ds9.si.edu/site/Home.html}
XSPEC \citep[vers.\ 12.9.0u,][]{arnaud96}.
}


\appendix

\section{Spectral fit in small-scale regions}

\begin{figure}
\epsscale{0.5}
\plotone{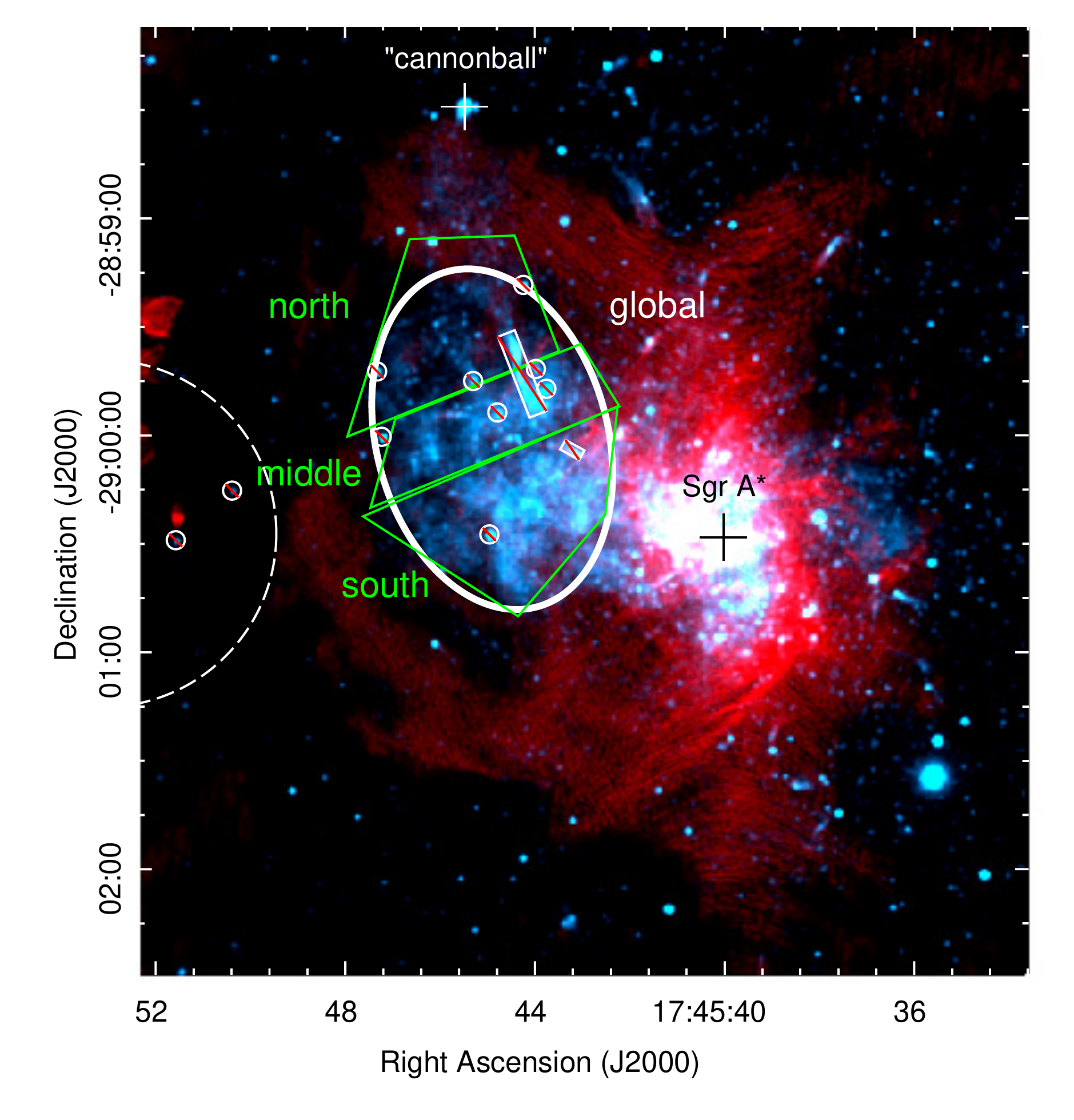}
\caption{
Composite image the same as Figure~1, overlaid with three additional regions
for spectral extraction.
The spectral fit results are shown in Table~\ref{tab:fit_3reg}.
}
\label{fig:x_c_3reg}
\end{figure}

In order to check if the abundance pattern significantly varies across the SNR and if it influences our conclusion on the origin of \snr, we extracted
spectra from three small-scale regions (``north,'' ``middle,'' and ``south'') as shown in 
Figure~\ref{fig:x_c_3reg}.
Similar to the global spectrum, the spectra from
the three regions can be described using a two-temperature model (see Table~\ref{tab:fit_3reg}).
The temperature of the cool component and 
the abundances of IMEs are consistent with the global spectrum, while
the absorption column density $N_{\rm H}$ and 
Fe abundances are increased toward the south.
The hot-component emission mainly arises from the regions ``middle''
and ``south,''
where the best-fit abundances are closest to those from the global
spectrum.

\begin{table}[!hb]
\footnotesize
\centering
\caption{Best-fit results for three individual regions  with 90\% uncertainties \label{tab:fit_3reg}}
\begin{tabular}{lccc}
\hline
\hline
Region & North  &  Middle  & South  \\
\hline
$\chi_\nu^2$/dof 
& 1.20/329
& 1.20/353
& 1.19/382 \\
$N_{\rm H} (10^{23}$~cm$^{-2})$  
& $1.88\pm 1.0$
& $1.97\pm 0.07$
& $2.30\pm 0.06$ \\
$kT_{\rm c}$ (keV) 
& $1.12\pm 0.07$
& $1.14\pm 0.05$
& $1.21\pm 0.05$ \\
$kT_{\rm h}$ (keV) 
& $7.4^{+2.5}_{-1.5}$
& $4.5\pm 0.4$
& $3.9\pm 0.3$ \\
$norm_{\rm c} (\times 10^{-2})$
& $1.1\pm 0.3$
& $1.5\pm 0.2$
& $2.4\pm 0.2$ \\
$norm_{\rm h} (\times 10^{-3})$  
& $0.4\pm 0.1$
& $5.9\pm 0.1$
& $1.1\pm 0.2$ 
\\
\hline
S
& $1.8\pm 0.4$
& $1.4\pm 0.3$
& $1.2\pm 0.3$ 
\\
Ar  
& $1.9\pm 0.4$
& $1.5\pm 0.3$
& $1.5\pm 0.3$ 
\\
Ca 
& $1.6\pm 0.3$
& $1.9\pm 0.2$
& $1.7\pm 0.2$ 
\\
Cr 
& $4.3\pm 3.2$
& $7.0\pm 2.0$
& $5.1\pm 1.3$ 
\\
Mn 
& $15 \pm 9$
& $12.6\pm 4.7$ & $18.7\pm 3.5$ 
\\
Fe 
& $2.4\pm 0.5$
& $5.3\pm0.6$ & $8.3\pm 0.6$ 
\\
Ni 
& $19\pm 12$
& $13.6\pm 5.5$ & $20.3\pm 4.9$ 
\\
\hline
\end{tabular}
\end{table} 

The thermonuclear origin of \snr\ is still favored
for these smaller regions, as indicated by the 
large Mn/Fe abundance ratios (2--6) and  the high IGE/IME ratios.
PTD remains the best explanation
for the individual regions.
The Fe abundance in the northern region is lower, indicating less
Fe or a stronger mixing with the ambient medium.
The other IGE abundances in the northern region have large uncertainties.

Moreover, we note that the single-component models $vvrnei$ and $bvvrnei$ for 
the global spectrum in the 5--8~keV band 
(see Table~\ref{tab:fit_3reg}) also suggest
high Mn/Fe and Cr/Fe ratios typical for the PTD models,
although the single-component model does not give a good fit
as the two-temperature model.

\section{Gas parameters and SNR properties} \label{sec:snr_properties}

The best-fit two-temperature model implies that
the plasma in \snr\ is a mixture of cool and hot gas.
Assuming that the X-ray bremsstrahlung emission is
from the plasma with near-solar metallicity, the gas densities 
in the cool and hot phases are derived as  $n_{\rm H}^{\rm c}=
11.2\pm 1.5 f^{-1/2}$~cm$^{-3}$ and $n_{\rm H}^{\rm h}=3.1\pm 0.7 f^{-1/2}$~cm$^{-3}$, respectively, where $f$ is the filling factor
of the observed X-ray-emitting gas in the whole volume.
The densities are calculated using the best-fit emission measures (proportional to the parameter $norm$) and an
assumed a prolate ellipsoid geometry of the X-ray-emitting plasma 
(short half-axis of $32''$ and long half-axis of $48''$, as shown
in Figure~\ref{fig:x_c_3reg}).
The cool and hot gas phases are considered to fill
$f$ of the whole volume and in pressure equilibrium, $n_{\rm H}^{\rm c} kT_{\rm c}$=$n_{\rm H}^{\rm h} kT_{\rm h}$.
This gives the filling factor of the cool phase
as $f_{\rm c}=0.65\pm 0.15f$.

We also obtain the gas mass 
to be $3.6\pm 0.4 f^{1/2}~M_\odot$ in the cool phase
and $0.55\pm 0.11 f^{1/2}~M_\odot$ in the hot phase.
Therefore, assuming that the ejecta and ISM 
are well mixed, the total mass of the X-ray-emitting
gas is $4.2\pm0.4 f^{1/2}~M_\odot$.
Then the total observed IGE masses are obtained 
as 
$M$(Cr)$= 3.2 \pm 0.8\times 10^{-4} f^{1/2}~M_\odot$,
$M$(Mn)$= 7.3\pm 1.4\times 10^{-4} f^{1/2}~M_\odot$,
$M$(Fe)$= 3.0 \pm 0.4\times 10^{-2} f^{1/2}~M_\odot$, and
$M$(Ni)$= 4.8\pm 1.2\times 10^{-3} f^{1/2}~M_\odot$.
These values are around one order of magnitude smaller than the values predicted in the best-fit PTD5 model with centered flame structures.
Nevertheless, we note the possibility that the X-ray emission might be partly from the pure SN ejecta \citep{park05}.
In this case,  the metal masses could be greatly enhanced with the same X-ray emission measure.

Pure-Fe ejecta has been found in the young SNR Cassiopeia~A \citep{hwang12}.
It has also been speculated that pure metal exists in the middle-age SNR W49B \citep{vink12, zhou18a,greco20}.
Similar to \snr, W49B is also a Fe-rich, mixed-morphology SNR interacting with dense ambient medium 
\citep[e.g.,][]{keohane07,miceli10,zhu14}.
In the pure ejecta case, the Bremsstrahlung emission 
measure is expressed as $EM = \sum n_e n_i Z_i^2 V$, where $n_e$, $n_i$, $Z_i$ and $V$
are the electron density, ion density, charge on the ion, and the ejecta volume.
Consequently, 
the IGE masses could be much larger than the values derived above.
The degeneracy between the best-fit abundances and emission measure 
leads to big uncertainties
in mass estimates.
Recently, \cite{greco20} pointed out that a bright radiative recombining continuum shows up when the plasma is made of pure-metal ejecta.
This introduces further difficulties in distinguishing recombining plasma and pure ejecta plasma
using the current CCD detectors.

Assuming that most Fe ions are He-like 
and the
ejecta have a filling factor $f$, we obtained
the Fe densities in the cool and hot components
to be $0.10\pm 0.01 f^{-1/2}$ cm$^{-3}$ and
$0.029\pm 0.006 f^{-1/2}$ cm$^{-3}$, respectively.
The total Fe mass in this extreme case is thus $1.6\pm 0.3 f^{1/2} M_\odot$.
This is an upper limit of Fe mass, as
the filling factor $f$ of pure ejecta cannot
be larger than 1 and the ejecta should be mixed with 
some shocked ISM.
Moreover, the Fe mass estimated here highly depends on the assumption of the three-dimensional morphology of the X-ray-emitting region.
We assumed a larger volume, resulting in a larger Fe mass than that
derived in \cite{park05}.
Therefore, the observed IGE masses have large uncertainties, although their abundance ratios are not affected by the mixing problem 
for \snr.

One explanation for the two-temperature
gas is that the cool and hot components
correspond to the gas shocked by the blast waves and reverse shock, respectively.
However, it is hard to explain why the abundances of the gas shocked by the blast wave are also rich in IGEs.
In an alternative explanation, the cool component corresponds
to the denser clumps, while the hot component is from the intercloud medium.
Both components are heated by the reverse shock.
As the X-ray photons below 2~keV are strongly absorbed, we do not know if
there is a third component colder than 1~keV. 

\snr\ is among the smallest SNRs, with a size 
of 5.5~pc $\times 7.8$~pc \citep{zhao16}, around half the size 
of W49B with an age of 5--6~kyr \citep{zhou18a,sun20}.
This indicates that \snr\ is either young, evolving in a 
dense medium, or has smaller explosion energy.
Using Sedov-Taylor self-similar solution
\citep{sedov59,taylor50, ostriker88},
we find the explosion energy is 
$E_0=(1/4\xi)(1.4 n_0 m_{\rm H}) R_{\rm s}^5 t_{\rm s}^{-2}
\sim 3.2\times 10^{50} (n_0/10~{\rm cm}^{-3})(R_{\rm s}/3.3~{\rm pc})^5 (t/ {\rm 2~kyr})^{-2}$~erg,
where $\xi=2.026$ and $n_0$ is the mean ambient density,
$R_{\rm s}$ is the radius of the SNR (the major and minor axes of the SNR
are 2.75~pc and 3.9~pc, respectively), and $t$ is the SNR age.

There is no consensus on the age of \snr.
Hydrodynamic simulations favored an early stage of the SNR \citep[$\lesssim 2$~kyr,][]{rockefeller05,fryer06}. 
In contrast, a few earlier X-ray studies suggested a large age for \snr,
in order to explain the (near-)CIE state of the plasma or to establish a presumed connection with the ``cannonball'' neutron star  \citep[5000--$10^4$~yr,][]{maeda02,sakano04}.
We note that the large ionization timescale of 
\snr's plasma does not necessarily support a large age.
The ionization timescale $\tau$ of X-ray plasma has been used to infer the SNR age $t=\tau/\nel$,
based
on an assumption of a constant electron density and simplest
ionization history.
If the spectrum is not dominated by the pure ejecta,
our spectral fit gives a 1$\sigma$ shock age of 2--$3f^{1/2}$~kyr for the cool component,
consistent with the young stage of \snr.
The hot component is in CIE, but the origin of this CIE state cannot
be simply explained with a large shock age.
Unlike the cool component, the electron temperature 
of the hot component varies by a factor of two across the SNR, as shown in previous studies \citep{park05,sakano04} and in Table~\ref{tab:fit_3reg}.
Such a large temperature gradient has also been found in the hot component of W49B, 
which has overionized plasma \citep[0.6--2.2~keV,][]{miceli10,lopez13b,zhou18a,yamaguchi18}. 
The variation of electron temperature and CIE state could result from  rapid cooling 
processes such as 
thermal conduction with cold gas or adiabatic  cooling.
Moreover, a high ionization stage of ions could be caused by the X-ray photoionization from past flares of Sgr A* \citep{one19}.

\snr\  is impacting a dense molecular shell
in the east
\citep{mezger96, yusef-zadeh00}, 
which was unlikely to be swept up by the SNR itself, as it would require a too-energetic SN explosion \citep[$E_0\sim 4\times 10^{52}$~erg][]{mezger89}.
We have shown that \snr\ originates from a
PTD WD explosion, which cannot have an ultrahigh explosion energy.
The density of the X-ray-emitting gas is $\lesssim 10$ cm$^{-3} f^{-1/2}$, a few orders
of magnitude smaller than that of a dense molecular cloud.
Therefore, our study supports that \snr\ was evolving in a relatively low-density medium 
until its shocks impacted 
the preexisting molecular shell \citep{serabyn92}.
A possible source that shaped the molecular 
shell is the winds of irrelevant massive 
stars in the SNR interior.
Another source of the winds could be strong 
accretion outflows from the progenitor
WD binary system prior to the SN explosion \citep{hachisu96}.
Wind-blown molecular bubbles have been reported for 
Type Ia SNR Tycho \citep{zhou16a,chen17} and N103B \citep{sano18}. Dense wind bubbles have also been found in Type Ia SNR candidates RCW 86 \citep{williams11} and W49B \citep{keohane07,zhou18a}.

\section{Numerical simulations of Type Ia SN models} \label{sec:simulation}

We have the formalism presented in detailed parameter surveys for
two-dimensional Type Ia SNe of near-Chandrasekhar-mass WD models
with or without deflagration-detonation transition \citep{leung2015a,leung2018Chand,leung20Iax} 
and for sub-Chandrasekhar-mass WD \citep{leung20subchand}.
We use the three-step nuclear
reactions \citep{Townsley2007,Calder2007} with parameterized timescales for
the carbon deflagration and detonation.
The chemical composition is described by a seven-isotope network \citep{Timmes20007iso}, including 
$^{4}$He, $^{12}$C, $^{16}$O, $^{20}$Ne, $^{24}$Mg, $^{28}$Si, and $^{56}$Ni.
To capture the deflagration and detonation, we use the level-set method \citep{Osher1988,Reinecke1999}.
The prescription of the sub-grid-scale turbulence model \citep{Niemeyer1995,Schmidt2006} is specific for the turbulent deflagration with a specific turbulent flame formula \citep{Hicks2015}. 
To determine whether the DDT occurs, 
we use the criteria by comparing the local size of the eddy motion with the flame width \citep{Golombek2005}.
For the sub-Chandrasekhar-mass models, where no deflagration takes place, we check 
the trigger of the carbon detonation by using the local density and temperature \citep{fink14}.
We use the tracer particle scheme \citep{Travaglio2004,Seitenzahl2010}
to record the thermodynamical history of the star. The tracers are designed to be massless and 
are passively advected by the fluid motion. They record the density and temperature
experienced according to the local quantities from the Eulerian meshes. 
To compute the nucleosynthesis yield, we choose a large
495-isotope network \citep{Timmes1999torch} which includes isotopes from 
$^{1}$H to $^{91}$Tc, and we apply this network to the thermodynamical trajectories 
of all the tracer particles. 

For the near-Chandrasekhar-mass WD models,
the simulations are done by first setting
up a C+O WD with a given central density $\rho_c$ (metallicity $Z=Z_\odot$ in this work);
this corresponds to a specific WD mass $M$. Then we put in the initial nuclear 
deflagration by hand at the center or at off-center. We allow the deflagration to propagate 
and interact with the fluid motion. For the models with DDT, we further set the code to check if
the flame front satisfies the transition criteria. If this is achieved, we put in by hand 
carbon detonation bubbles and allow them to propagate independently with the carbon deflagration waves. 
We carry out the simulation until 
the star develops into homologous expansion and becomes sufficiently cold such that 
no major exothermic nuclear reactions continue.

For the sub-Chandrasekhar-mass WD models, the simulations are done by first 
setting a C+O WD with a helium envelope. The central density 
and the transition density from the C$+$O core to the He envelope are chosen such 
that the total mass $M$ and the He-envelope $M_{\rm He}$ mass are as required. 
After that, we put in the initial detonation by hand in the He envelope
and start the simulation. The local thermodynamical condition in the C+O core
is checked to see if the second detonation can be triggered by the 
shock wave interactions generated by the He detonation. When the condition
is met, a carbon detonation bubble is put in by hand, and we let the 
carbon detonation and helium detonation propagate independently until 
the star is totally disrupted and reaches homologous expansion.

Table~\ref{tab:tnmodel} shows the basic 
parameters of the thermonuclear SN models
used in Figure~\ref{fig:progenitor}, where the original names of the models are listed.
We also summarize the nucleosynthesis yields of the thermonuclear models  in Table~\ref{tab:ddt_dd} and Table \ref{tab:ptd}.

\begin{table}[!hb]
\footnotesize
\centering
\caption{Basic parameters and model conversion
for thermonuclear SN models. 
}
\label{tab:tnmodel}
\begin{threeparttable}
\begin{tabular}{l|c|c|c|c|c|c}
\hline
\hline
Model & Explosion Type & Ejecta Mass & Remnant Mass & $^{56}$Ni & $^{44}$Ti & Model in Reference \\ 
& & ($M_\odot$) &  ($M_\odot$) & ($M_\odot$) & ($M_\odot$) & \\ 
\hline
DDT2   & DDT & 1.35 & & 0.70 & 3.28e-5 & 200-1-c3-1 \\
DDT3   &     & 1.37 &  & 0.63 & 2.47e-5 & 300-1-c3-1 \\
DDT5   &     & 1.38 &  & 0.60 & 2.29e-5 & 500-1-c3-1 \\
DDT5.5 &     & 1.38 &  & 0.75 & 2.76e-5 & 550-1-c3-1 \\ \hline
Bubble ($1.1 \Msun$) & DD  & 1.10 &  & 0.61 & 5.14e-4 & 110-100-2-50 \\
Ring ($1.1 \Msun$)  &     & 1.10 &  & 0.68 & 5.99e-4 & 110-050-2-B50 \\
Sphere ($1.0 \Msun$) &     & 1.00 &  &0.60 & 2.64e-4 & 100-050-2-S50 \\ 
Sphere ($0.9 \Msun$) &     & 0.90 &  &0.0155 & 3.27e-5 & 090-050-2-S50 \\ \hline
PTD2 & PTD & 1.18 & 0.17 & 0.24 & 1.64e-6 & 200-135-1-c3-1 \\
PTD3   &     & 1.26 & 0.11& 0.34 & 1.79e-6 & 300-137-1-c3-1 \\
PTD5   &     & 1.29 & 0.09 & 0.32 & 1.75e-6 & 500-138-1-c3-1 \\
PTD5.5 &     & 1.30 & 0.08 & 0.31 & 1.51e-6 & 550-138-1-c3-1 \\
\hline
\end{tabular}
\begin{tablenotes}
\item
``DD,'' ``DDT,''  and ``PTD'' 
correspond to the double detonation model \citep{leung20subchand}, turbulent deflagration model with  
deflagration-detonation transition \citep{leung2018Chand},
and
pure turbulent deflagration model, respectively \citep{leung20Iax}.
\end{tablenotes}
\end{threeparttable}
\end{table}

\begin{table}
\footnotesize
\caption{
Nucleosynthesis yields in unit of $M_\odot$ for the DDT and DD models 
\citep[Type Ia,][]{leung2018Chand,leung20subchand}.
}
\label{tab:ddt_dd}
\begin{center}
\begin{tabular}{l|c|c|c|c|c|c|c}
\hline
\hline
Element & DDT2 & DDT3 & DDT5 & DDT5.5 & Bubble & Ring & Sphere \\ \hline
C & $1.4 \times 10^{-3}$ & $1.7 \times 10^{-3}$ & $5.82 \times 10^{-4}$ & $4.38 \times 10^{-6}$ & $3.35 \times 10^{-3}$ & $4.2 \times 10^{-3}$ & $1.15 \times 10^{-3}$ \\
N & $1.72 \times 10^{-10}$ & $2.34 \times 10^{-10}$ & $5.2 \times 10^{-10}$ & $5.60 \times 10^{-11}$ & $3.90 \times 10^{-8}$ & $1.18 \times 10^{-7}$ & $1.84 \times 10^{-8}$ \\
O & $3.83 \times 10^{-2}$ & $5.69 \times 10^{-2}$ & $4.90 \times 10^{-2}$ & $1.29 \times 10^{-2}$ & $1.17 \times 10^{-1}$ & $1.2 \times 10^{-1}$ & $6.64 \times 10^{-2}$ \\
F & $3.18 \times 10^{-14}$ & $1.38 \times 10^{-13}$ & $8.34 \times 10^{-13}$ & $1.59 \times 10^{-16}$ & $3.28 \times 10^{-11}$ & $6.80 \times 10^{-11}$ & $2.39 \times 10^{-11}$ \\
Ne & $1.66 \times 10^{-4}$ & $1.81 \times 10^{-4}$ & $6.77 \times 10^{-4}$ & $2.15 \times 10^{-6}$ & $4.9 \times 10^{-3}$ & $4.81 \times 10^{-3}$ & $1.15 \times 10^{-3}$ \\
Na & $5.38 \times 10^{-7}$ & $8.9 \times 10^{-7}$ & $2.0 \times 10^{-6}$ & $4.57 \times 10^{-8}$ & $1.47 \times 10^{-5}$ & $2.10 \times 10^{-5}$ & $8.39 \times 10^{-6}$ \\
Mg & $8.54 \times 10^{-4}$ & $1.11 \times 10^{-3}$ & $1.24 \times 10^{-3}$ & $3.42 \times 10^{-4}$ & $8.36 \times 10^{-3}$ & $8.75 \times 10^{-3}$ & $1.32 \times 10^{-3}$ \\
Al & $7.17 \times 10^{-5}$ & $9.14 \times 10^{-5}$ & $9.18 \times 10^{-5}$ & $2.87 \times 10^{-5}$ & $6.50 \times 10^{-4}$ & $7.17 \times 10^{-4}$ & $1.14 \times 10^{-4}$ \\
Si & $2.21 \times 10^{-1}$ & $2.35 \times 10^{-1}$ & $2.9 \times 10^{-1}$ & $1.6 \times 10^{-1}$ & $1.37 \times 10^{-1}$ & $1.13 \times 10^{-1}$ & $1.25 \times 10^{-1}$ \\
P & $1.33 \times 10^{-4}$ & $1.92 \times 10^{-4}$ & $1.65 \times 10^{-4}$ & $4.17 \times 10^{-5}$ & $3.97 \times 10^{-4}$ & $3.54 \times 10^{-4}$ & $1.53 \times 10^{-4}$ \\
S & $1.20 \times 10^{-1}$ & $1.25 \times 10^{-1}$ & $1.11 \times 10^{-1}$ & $6.64 \times 10^{-2}$ & $6.38 \times 10^{-2}$ & $5.13 \times 10^{-2}$ & $6.77 \times 10^{-2}$ \\
Cl & $1.50 \times 10^{-4}$ & $2.4 \times 10^{-4}$ & $1.82 \times 10^{-4}$ & $3.78 \times 10^{-5}$ & $1.84 \times 10^{-4}$ & $1.54 \times 10^{-4}$ & $1.29 \times 10^{-4}$ \\
Ar & $2.32 \times 10^{-2}$ & $2.41 \times 10^{-2}$ & $2.14 \times 10^{-2}$ & $1.47 \times 10^{-2}$ & $1.10 \times 10^{-2}$ & $8.57 \times 10^{-3}$ & $1.29 \times 10^{-2}$ \\
K & $1.44 \times 10^{-4}$ & $1.89 \times 10^{-4}$ & $1.75 \times 10^{-4}$ & $3.54 \times 10^{-5}$ & $1.15 \times 10^{-4}$ & $9.9 \times 10^{-5}$ & $9.82 \times 10^{-5}$ \\
Ca & $1.83 \times 10^{-2}$ & $1.80 \times 10^{-2}$ & $1.60 \times 10^{-2}$ & $1.43 \times 10^{-2}$ & $9.7 \times 10^{-3}$ & $7.52 \times 10^{-3}$ & $1.7 \times 10^{-2}$ \\
Sc & $5.21 \times 10^{-7}$ & $6.5 \times 10^{-7}$ & $5.71 \times 10^{-7}$ & $2.18 \times 10^{-7}$ & $1.67 \times 10^{-6}$ & $7.81 \times 10^{-7}$ & $3.77 \times 10^{-7}$ \\
Ti & $4.50 \times 10^{-4}$ & $4.9 \times 10^{-4}$ & $9.16 \times 10^{-4}$ & $1.53 \times 10^{-3}$ & $5.5 \times 10^{-3}$ & $2.72 \times 10^{-3}$ & $7.64 \times 10^{-4}$ \\
V & $8.47 \times 10^{-5}$ & $9.47 \times 10^{-5}$ & $2.88 \times 10^{-4}$ & $3.59 \times 10^{-4}$ & $5.90 \times 10^{-4}$ & $2.81 \times 10^{-4}$ & $7.62 \times 10^{-5}$ \\
Cr & $7.19 \times 10^{-3}$ & $9.58 \times 10^{-3}$ & $2.22 \times 10^{-2}$ & $2.78 \times 10^{-2}$ & $1.68 \times 10^{-2}$ & $3.7 \times 10^{-3}$ & $3.29 \times 10^{-3}$ \\
Mn & $8.70 \times 10^{-3}$ & $1.3 \times 10^{-2}$ & $1.34 \times 10^{-2}$ & $1.43 \times 10^{-2}$ & $2.28 \times 10^{-3}$ & $9.15 \times 10^{-4}$ & $1.32 \times 10^{-3}$ \\
Fe & $8.23 \times 10^{-1}$ & $7.99 \times 10^{-1}$ & $8.37 \times 10^{-1}$ & $10.7 \times 10^{-1}$ & $6.40 \times 10^{-1}$ & $7.8 \times 10^{-1}$ & $6.25 \times 10^{-1}$ \\
Co & $8.22 \times 10^{-4}$ & $8.85 \times 10^{-4}$ & $1.25 \times 10^{-3}$ & $1.47 \times 10^{-3}$ & $7.19 \times 10^{-4}$ & $6.17 \times 10^{-4}$ & $5.27 \times 10^{-4}$ \\
Ni & $7.31 \times 10^{-2}$ & $7.70 \times 10^{-2}$ & $8.69 \times 10^{-2}$ & $9.84 \times 10^{-2}$ & $4.37 \times 10^{-2}$ & $3.91 \times 10^{-2}$ & $3.31 \times 10^{-2}$ \\
Cu & $5.83 \times 10^{-6}$ & $4.12 \times 10^{-6}$ & $7.29 \times 10^{-6}$ & $1.7 \times 10^{-5}$ & $2.32 \times 10^{-5}$ & $5.73 \times 10^{-6}$ & $1.1 \times 10^{-5}$ \\
Zn & $6.99 \times 10^{-5}$ & $4.53 \times 10^{-5}$ & $4.76 \times 10^{-5}$ & $6.87 \times 10^{-5}$ & $1.79 \times 10^{-4}$ & $3.62 \times 10^{-5}$ & $5.19 \times 10^{-5}$ \\
 \hline 
\end{tabular}
\end{center}
\end{table}

\begin{table}
\footnotesize
\begin{center}
\caption{
Nucleosynshesis yields in unit of $M_\odot$ for the PTD models \citep[Type Iax,][]{leung20Iax}
}
\label{tab:ptd}
\begin{tabular}{l|c|c|c|c}
\hline
\hline
Element & PTD2 & PTD3 & PTD5 & PTD5.5
\\
\hline 
C & $3.81 \times 10^{-1}$ & $3.21 \times 10^{-1}$ & $3.14 \times 10^{-1}$ & $3.14 \times 10^{-1}$ \\ 
 N  & $5.29 \times 10^{-9}$ & $7.29 \times 10^{-9}$ & $6.67 \times 10^{-9}$ & $6.15 \times 10^{-9}$ \\
 O & $4.8 \times 10^{-1}$ & $3.54 \times 10^{-1}$ & $3.44 \times 10^{-1}$ & $3.40 \times 10^{-1}$ \\
 F  & $1.52 \times 10^{-11}$ & $1.81 \times 10^{-11}$ & $1.75 \times 10^{-11}$ & $1.65 \times 10^{-11}$ \\
 Ne  & $1.86 \times 10^{-2}$ & $1.63 \times 10^{-2}$ & $1.59 \times 10^{-2}$ & $1.59 \times 10^{-2}$  \\
 Na & $9.76 \times 10^{-6}$ & $1.5 \times 10^{-5}$ & $1.5 \times 10^{-5}$ & $1.2 \times 10^{-5}$ \\
 Mg & $3.63 \times 10^{-3}$ & $4.5 \times 10^{-3}$ & $3.80 \times 10^{-3}$ & $3.49 \times 10^{-3}$ \\
 Al & $2.67 \times 10^{-4}$ & $3.1 \times 10^{-4}$ & $2.78 \times 10^{-4}$ & $2.59 \times 10^{-4}$ \\
 Si & $3.72 \times 10^{-2}$ & $4.10 \times 10^{-2}$ & $4.45 \times 10^{-2}$ & $3.69 \times 10^{-2}$ \\
 P  & $1.6 \times 10^{-4}$ & $1.27 \times 10^{-4}$ & $1.17 \times 10^{-4}$ & $1.0 \times 10^{-4}$ \\
 S  & $1.67 \times 10^{-2}$ & $1.82 \times 10^{-2}$ & $1.99 \times 10^{-2}$ & $1.64 \times 10^{-2}$ \\
 Cl  & $3.81 \times 10^{-5}$ & $4.40 \times 10^{-5}$ & $4.24 \times 10^{-5}$ & $3.71 \times 10^{-5}$ \\
 Ar & $2.83 \times 10^{-3}$ & $3.10 \times 10^{-3}$ & $3.35 \times 10^{-3}$ & $2.75 \times 10^{-3}$ \\
 K  & $1.75 \times 10^{-5}$ & $1.93 \times 10^{-5}$ & $2.11 \times 10^{-5}$ & $1.59 \times 10^{-5}$ \\
 Ca & $2.13 \times 10^{-3}$ & $2.35 \times 10^{-3}$ & $2.48 \times 10^{-3}$ & $2.6 \times 10^{-3}$ \\
 Sc & $3.64 \times 10^{-8}$ & $4.41 \times 10^{-8}$ & $4.68 \times 10^{-8}$ & $3.63 \times 10^{-8}$ \\
 Ti& $4.31 \times 10^{-5}$ & $5.60 \times 10^{-5}$ & $5.73 \times 10^{-5}$ & $1.4 \times 10^{-4}$ \\
 V & $1.45 \times 10^{-5}$ & $2.73 \times 10^{-5}$ & $3.60 \times 10^{-5}$ & $6.88 \times 10^{-5}$ \\
 Cr & $1.38 \times 10^{-3}$ & $2.61 \times 10^{-3}$ & $4.14 \times 10^{-3}$ & $7.72 \times 10^{-3}$ \\
 Mn & $3.29 \times 10^{-3}$ & $7.18 \times 10^{-3}$ & $9.3 \times 10^{-3}$ & $1.1 \times 10^{-2}$ \\
 Fe & $2.78 \times 10^{-1}$ & $4.36 \times 10^{-1}$ & $4.65 \times 10^{-1}$ & $4.83 \times 10^{-1}$ \\
 Co & $1.19 \times 10^{-4}$ & $4.5 \times 10^{-4}$ & $5.85 \times 10^{-4}$ & $7.6 \times 10^{-4}$ \\
 Ni & $2.45 \times 10^{-2}$ & $5.5 \times 10^{-2}$ & $6.5 \times 10^{-2}$ & $6.48 \times 10^{-2}$ \\
 Cu  & $1.57 \times 10^{-7}$ & $3.72 \times 10^{-7}$ & $3.28 \times 10^{-7}$ & $8.78 \times 10^{-7}$ \\
 Zn & $1.17 \times 10^{-6}$ & $2.63 \times 10^{-6}$ & $1.76 \times 10^{-6}$ & $1.62 \times 10^{-6}$ \\
 \hline
 \end{tabular}
\end{center}
\end{table}

\end{document}